\documentclass[prd,tightenlines,nopacs,nofootinbib,eqsecnum,amsfonts,amssymb,amsmath,twocolumn]{revtex4}
\pdfoutput=1
\usepackage{bm}
\usepackage{graphicx}

\newcommand{\etal}{{\it et al.}}
\newcommand{\dd}{{\rm d}}
\newcommand{\bx}{\bm{x}}
\newcommand{\bk}{\bm{k}}
\newcommand{\bn}{\bm{n}}

\def\be{\begin{equation}}
\def\ee{\end{equation}}
\def\bea{\begin{eqnarray}}
\def\eea{\end{eqnarray}}
\renewcommand{\mathbf}[1]{\bm{#1}}
\newcommand{\sfrac}[2]{{\textstyle\frac{#1}{#2}}}
\begin{document}

\title{Non-linear relativistic contributions to the cosmological weak-lensing convergence}

\author{Sambatra Andrianomena$^1$, Chris Clarkson$^1$, Prina Patel$^{1,2}$, Obinna Umeh$^{1,2}$ and Jean-Philippe Uzan$^3$}
\email{andrianomena@gmail.com, chris.clarkson@gmail.com, prina83@gmail.com, umeobinna@gmail.com, uzan@iap.fr}
\affiliation{$1$-- Astrophysics, Cosmology and Gravity Centre, and, Department of Mathematics and Applied Mathematics, University of Cape Town, Rondebosch 7701, South Africa\\
$2$-- Physics Department, University of the Western Cape,
Cape Town 7535, South Africa\\
$3$-- Institut d'Astrophysique de Paris, UMR-7095 du CNRS, Universit\'e Pierre et Marie Curie, 98 bis bd Arago, 75014 Paris, France,\\
and, Sorbonne Universit\'es, Institut Lagrange de Paris, 98 bis bd Arago, 75014 Paris, France.}

\begin{abstract}
Relativistic contributions to the dynamics of structure formation come in a variety of forms, and can potentially give corrections to the standard picture on typical scales of 100\,Mpc. These corrections cannot be obtained by Newtonian numerical simulations, so it is important to accurately estimate the magnitude of these relativistic effects.  Density fluctuations couple to produce a background of gravitational waves, which is larger than any primordial background. A similar interaction produces a much larger spectrum of vector modes which represent the frame-dragging rotation of spacetime.  These can change the metric at the percent level in the concordance model at scales below the equality scale. Vector modes modify the lensing of background galaxies by  large-scale structure. This gives in principle the exciting possibility of measuring relativistic frame dragging effects on cosmological scales. The effects of the non-linear tensor and vector modes on the cosmic convergence are computed and compared to first-order lensing contributions from density fluctuations, Doppler lensing, and smaller Sachs-Wolfe effects. The lensing from gravitational waves is negligible so we concentrate on the vector modes. We show the relative importance of this for future surveys such as Euclid and SKA. We find that these non-linear effects only marginally affect the overall weak lensing signal so they can safely be neglected in most analyses, though are still much larger than the linear Sachs-Wolfe terms. The second-order vector contribution can dominate the first-order Doppler lensing term at moderate redshifts and are actually more important for survey geometries like the SKA.

\end{abstract}
 \maketitle

\section{Introduction}\label{sec1}

Relativistic corrections to the standard model of cosmology come in a variety of forms, from the altering the dynamics of structure formation to the various effects associated to the interpretation of observations, in particular modifying the propagation of light. 

There has been considerable debate as to the importance and amplitude of these effects on the dynamics of the expansion of the universe and the growth of large scale structure (see, e.g., Ref.~\cite{Clarkson:2011zq} for an overview), and the amplitude and importance of these dynamical effects are still actively debated~\cite{buchertetal,nobackreac,backsig}. Though subdominant for linear structure formation, relativistic corrections are a generic prediction of General Relativity and are inevitable at a non-linear level through mode-mode coupling. The scalar gravitational potential induces rotational frame-dragging modes in spacetime (so-called vector modes) as well as gravitational waves (tensor modes). Neither of these have counterparts in Newtonian gravity as they both induce a non-zero magnetic Weyl curvature which is absent in Newtonian gravity and difficult to take into account in N-body numerical simulations~\cite{Bruni:2013mua,Adamek:2013wja}. They therefore serve as an important tool in understanding purely relativistic aspects of structure formation and its observational consequences, as they set a lower limit on the amplitude of relativistic corrections. 

On top of dynamical corrections, relativistic effects also induce corrections to the propagation of light since it probes the complete spacetime geometry. This can alter the interpretation of cosmological observations at a level that cannot be neglected in an era of ``precision cosmology''. Provided one works within perturbation theory, the amplitude of these effects is computable and completely fixed once the normalisation of the scalar power spectrum, at the linear level, is determined. For instance, some relativistic effects have been taken into account on the cosmic microwave background~\cite{pub} and shown to be below the constraints on non-Gaussianity derived by \textit{Planck}~\cite{planckXXIX}, but nevertheless in principle detectable on small angular scales, in particular through spectral distortions~\cite{spectral}. This article focuses on the effect of relativistic corrections on weak lensing observations, focusing mainly on the induced vector mode background. Weak gravitational lensing by the large-scale structure of the Universe has now become a major tool of cosmology~\cite{revuesWL}, used to study questions ranging from the distribution of dark matter to tests of general relativity~\cite{testGR}.

The propagation of light in an inhomogeneous universe gives rise to both distortion and magnification induced by gravitational lensing. The effect of non-linear corrections on the Hubble diagram have been considered~\cite{BenDayan:2012wi,cemuu,Umeh:2012pn,Umeh:2014ana} and shown to be non-negligible given the accuracy of contemporary observations~\cite{Fleury:2013sna,Fleury:2013uqa,voidslensing,BenDayan:2012ct,BenDayan:2013gc}. Previous works considered the contributions of the vector metric perturbations to the shear and magnification using standard rulers~\cite{Schmidt:2012mg,Jeong:2013ds}. In this article we consider the effect on the weak lensing convergence of non-linear effects that induce the existence of a vector and tensor modes background. We compare this to the various contributions to the convergence at first-order~-- the usual integral of the density contrast along the line of sight~\cite{revuesWL}, the contribution from the Doppler effect which is dominant at low redshifts and large scales~\cite{Bonvin:2008ni,voidslensing,Bacon:2014uja}, the Integrated Sachs-Wolfe (ISW) and Sachs-Wolfe (SW) terms which are relatively small and mainly neglected when computing cosmic convergence.

The induced background of gravitational waves from scalar-scalar coupling was presented in Ref.~\cite{Ananda:2006af} during the radiation era, and its present-day spectrum calculated in Ref.~\cite{Baumann:2007zm}, with shear lensing effects studied in Ref.~\cite{Sarkar:2008ii}, all following the pioneering analysis of Ref.~\cite{Mollerach:2003nq}. Surprisingly it was found that the induced gravitational wave background is significantly larger than any primordial background (even for a tensor-scalar ratio $r\sim0.1$) on intermediate scales of $\sim$100\,Mpc, which is around the equality scale, though of course it is much smaller on small scales. Similarly, the induced vector mode background was presented in Refs.~\cite{Lu:2007cj,Lu:2008ju}, and again a spectrum was found that peaks on 100\,Mpc scales. Remarkably, however, it was found that the amplitude of the background of vector modes for the metric potential behaves on small scales with the same scaling as the gravitational potential, with nearly 1\% of its amplitude. While both of these induced degrees of freedom have little effect on the dynamics of structure formation (they cannot directly source the density fluctuation as it is a scalar degree of freedom) they can influence the gravitational lensing produced by large-scale structure. Is it significant, and could it be a new way to detect relativistic aspects of structure formation?

The effects of these contributions on weak lensing convergence predictions are computed in order to understand if they can either be detected or, in the worst case, bias the analysis of future weak lensing experiments, such as Euclid or SKA; {\em i.e.}, if the interpretation of the observation by assuming that the observed convergence corresponds to the convergence sourced by scalar modes only is an accurate enough assumption or whether some of these effects have to be included in the analysis. This article addresses this question and computes the effect of these two non-linear effects on weak lensing observations by considering second order vector and tensor background. We restrict our analysis to the direct contribution from the dynamically induced vector modes and the hypothesis that the Born approximation still holds. In principle, one needs also to take into account second order effects on the geodesic deviation equation~\cite{Seitz:1994xf, Cooray:2002mj,Dodelson:2005zj,Schaefer:2005up,Shapiro:2006em}, as fully described in Refs.~\cite{Bernardeau:2009bm,Bernardeau:2011tc}. The calculations by relaxing the Born approximation will induce small changes to the signal. However, there are of course a variety of other geometrical effects which may dominate the signal \cite{Obinna:2012,Obinna:2014} but our goal in this study is to investigate the convergence from dynamical effects only..

In Section~\ref{sec2} we describe the vector and gravity waves background induced by the non-linear dynamics and then, in Section~\ref{sec3}, the computation of the weak lensing power spectra, splitting the effects of the scalar, vector, tensor, Doppler, ISW and SW contributions in order to compare their magnitude. Since the contribution of the tensor modes remains negligible and both ISW and SW being relatively small, we focus in Section~\ref{sec4} on the vector and Doppler contribution, estimating their magnitude in surveys such as Euclid and SKA. Technical details are gathered in Appendices~\ref{app0}-\ref{appC}.

\section{Induced vector and gravitational wave backgrounds}\label{sec2}

Let us start by briefly reviewing the vector and gravitational wave backgrounds induced by structure formation. In the standard cosmological framework, the initial conditions set by inflation imposes that at the linear order only scalar perturbations, described in \S~\ref{sec2-1}, are significantly sourced. At second order, one cannot neglect the contributions from vector and tensor modes, that are respectively described in \S~\ref{sec2-2} and~\ref{sec2-3}.

We shall work in the Poisson (or Newtonian) gauge in which the metric can be expanded as
\begin{eqnarray}\label{metric}
\dd s^2 &=&a(\eta)^2\left[-(1 + 2\Phi)\dd \eta^2 + 2V_{i} \dd x^{i}\dd \eta \right.\nonumber\\ &&\left.
 + \left((1-2 \Psi)\gamma_{i j} + h_{ij}\right)\dd x^{i}\dd x^{j}\right],
\end{eqnarray}
where 
$a$ is the scale factor, $\eta$ the conformal time and $\gamma_{ij}$ is the spatial metric of the background. Latin indices run from 1 to 3. The scalar, vector and tensor perturbations are respectively described by $\Phi$ and $\Psi$, $V_i$ and $h_{ij}$ where $V_i$ is transverse ($D_i V^i=0$) and $h_{ij}$ is transverse and traceless ($h^i_i=0$ and $D_ih^i_j=0$) where $D_i$ is the covariant derivative associated with $\gamma_{ij}$.

\subsection{First order scalar perturbations}\label{sec2-1}

At late times, we can neglect the anisotropic stress of matter (mostly described by a pressure-free fluid on cosmological scales) and the spatial curvature (so that we assume that the spatial sections are Euclidean). 

It follows that the Einstein equations imply (from the traceless part of the $(ij)$ Einstein equations) $\Phi=\Psi$ (see e.g., Ref.~\cite{pubook} for a derivation of the following equations).  The peculiar velocity sourced by first order scalars is given in terms of the potential, from the $(0i)$ component of the Einstein equations, as
\be\label{j2.2}
v_i(\eta,\bm x)= -\frac{2a}{3\Omega_{\rm m}H_{0}^{2}}\partial_i(\Phi'+\mathcal{H}\Phi)
\ee
where the conformal Hubble rate is defined as $\mathcal{H}=a'/a$, a prime denoting a derivative with respect to $\eta$. It is related to the Hubble rate by ${\cal H}=aH$. In a $\Lambda$CDM model in which the late time dynamics is dominated by a pure cosmological constant and dark matter, it is  given by
\be\label{friedeq}
{\cal H}=H_0\sqrt{\frac{\Omega_{\rm m}}{a}+a^2\Omega_\Lambda},
\ee
where $\Omega_{\rm m}$ and $\Omega_\Lambda$ are the matter and cosmological constant density parameters evaluated today.

The matter density contrast $\delta$ can be obtained from the relativistic Poisson equation, that derives from the Einstein equations. It involves the scalar component of the peculiar velocity $v$ ($v_i=\partial_iv$)
\be\label{grp} 
\delta = \frac{2a}{3\Omega_{\rm m}H_{0}^{2}}\Delta\Phi +3\mathcal{H}v, 
\ee
where $\Delta=\nabla^2$ is the 3 dimensional Laplacian.
The evolution of the gravitational potential is then obtained from the spatial trace of the Einstein equations, combined with Eq.~(\ref{grp}), to give
\begin{equation}
 \Phi'' + 3{\cal H}(1+c_s^2)\Phi' +[2{\cal H}'+{\cal H}^2(1+3c_s^2)]\Phi -c_s^2\Delta\Phi=0,
 \nonumber
\end{equation}
as long as the anisotropic stress can be neglected. $c_s^2$ is the sound speed. For a pressureless fluid, such as matter on cosmological scales, $c_s^2=0$ so that the solution of this equation can be factorized as $\Phi(\eta,\bm x)=g(\eta)\Phi_i(\bm x)$. $\Phi_i(\bm x)$ (or equivalently  $\Phi_i(\bm k)$ in Fourier space) describes the initial conditions. The growth suppression factor $g(\eta)$ is determined from
\be
g''(\eta)+ 3 \mathcal{H} g'(\eta) +  a^2\Lambda g(\eta)= 0,
\ee
where Eq.~(\ref{friedeq}) has been used to evaluate the third term. $g$ describes the growth of the gravitational potential after decoupling. In general, one uses the linearity of the perturbation equations to decompose the gravitational potential in terms of a transfer function $T$ as $\Phi(\bk,\eta)=T(k,\eta)\Phi_i(\bk)$ in Fourier space, defining the Fourier modes by
\be\label{e.Fdec}
\Phi(\bx,\eta) = \int\frac{\rm d^3\bk}{(2\pi)^{3/2}}\Phi(\bk,\eta)\hbox{e}^{-{\rm i} \bk\cdot\bx}.
\ee
It follows that the scalar power spectrum is defined as
\begin{equation}\label{e.powerS}
 \langle\Phi(\bk,\eta)\Phi^*(\bk',\eta')\rangle=\frac{2\pi^2}{k^3}\mathcal{P}_\Phi(k,\eta,\eta')
 \delta^{(3)}(\bk-\bk')\,,
\end{equation}
where $\delta^{(3)}$ stands for the Dirac distribution.

The power spectrum today can be related to the initial power spectrum predicted from inflation. Assuming scale invariance (which is a good approximation for our analysis since secondary modes are quite insensitive to the spectral index), the inflationary power spectrum is characterized by its primordial power $\Delta_{\mathcal{R}}^{2}$, typically of order $\Delta_{\mathcal{R}}^{2} \approx 2.41 \times 10^{-9}$ at a scale
$k_{\rm CMB}=0.002~\mathrm{Mpc}^{-1}$~\cite{wmap5}. It follows that
\be
\mathcal{P}_{\Phi}(k)=\left( \frac{3 \Delta_\mathcal{R}}{5 g_{\infty}} \right )^2 g^2(\eta)T^2(k), 
\ee
where $g_\infty$ is chosen so that $g(\eta_0)=1$. In the following, we shall use the transfer function derived in Ref.~\cite{Eisenstein:1997ik} to model the linear transfer function, and we also use Halofit~\cite{Smith:2002dz} to estimate nonlinear small scale effects. Due to non-linear evolution, the growth suppression factor becomes scale dependent as
\begin{equation}\label{halo}
g_{\rm nl}(\chi,k) = (z+1)\sqrt{\frac{P_{\rm nl}(\chi,k)}{P(k)}}.
\end{equation}
We then use this growth suppression factor to account for the non-linearities. Since non-linear evolution occurs at small scales (large $k$), $g_{\rm nl}(\chi,k)$ behaves as the linear $g(\chi)$ which is $k$ independent on large scales ($k$ small). $P_{\rm nl}(\chi,k)$ and $P(k)$ are the non-linear matter power spectrum and today's linear matter power spectrum respectively.

\subsection{Second order vector contribution}\label{sec2-2}

At second order, vector modes are sourced from the mode coupling of order 1 scalar modes, $\mathcal{O}(\Phi^2)$. Assuming Euclidean spatial sections, the second order Einstein equations in Newtonian gauge~\cite{O2cp,Lu:2008ju} lead to the second order vector contribution
\be \label{new solution for S}
V_i={16a\over 3\Omega_{\rm m}H_0^2}\Delta^{-1} \left\{
\Delta\Phi\,\partial_i \left( \Phi'+{\cal H}\Phi\right)\right\}^V\,,
\ee
where $V$ denotes the vector contribution of the part inside the braces. Of course estimating the non-linear corrections to the vectors in this way neglects a variety of other effects which could be important, so we include this as a rough estimate only. The Fourier transform of the vector perturbation encodes the two orthogonal polarisations and is defined as
\begin{equation}\label{eFvec}
 V_i(\bm x,\eta) = \int\frac{\dd^3\bk}{(2\pi)^{3/2}}\sum_{\lambda=\pm}
 V_\lambda(\bk,\eta) e_i^\lambda(\bk)\, \hbox{e}^{{\rm i}\bk\cdot\bx}\ ,
\end{equation}
where the two vectors $\lbrace\mathbf{e}^+,\mathbf{e}^-\rbrace$ realize an orthonormal basis orthogonal to $\bk$ ({\em i.e.}, $\mathbf{e}^\lambda\cdot\mathbf{e}^{\lambda^\prime}=\delta^{\lambda\lambda^\prime}$, $\mathbf{e}^\lambda\cdot\bk=0$).
The power in each polarisation  is, thanks to spatial isotropy, the same and is defined in the same way as for the scalars for each polarisation. During the matter dominated era the vector contribution grows as $a^{1/2}$ which is the reason why it is not completely negligible today~\cite{Lu:2007cj,Lu:2008ju}. Their contribution peaks in power at the equality scale, and has the same spectrum as $\Phi$ below this scale, but with $\lesssim$1\% of the amplitude~\cite{Lu:2008ju}. The vector mode power spectrum we shall use in our analysis can be parameterized~\cite{Lu:2008ju} as
\be\label{e.powerV}
\mathcal{P}_V(k,\eta,\eta') = \left(\frac{2\Delta_\mathcal{R}}{5g_\infty\sqrt{\Omega_{\rm m}}H_0}\right)^4 \mathcal{V}(\eta)\mathcal{V}(\eta')
 k^2 \Pi(k)\,,
\ee
where
\be
\mathcal{V}(\eta)=3a(\eta)g(\eta)[g'(\eta)+\mathcal{H}(\eta)g(\eta)]
\ee
governs the growth of the vector power spectrum, and  $\Pi(k)$ is a convolution integral of order unity (see Eq. (C7) of Ref.~\cite{Lu:2008ju} for its explicit expression). The amplitude of the vectors decays on scales smaller than the equality scale, $k> k_{\rm eq}\approx0.073\,\Omega_{\rm m}h^2\text{Mpc}^{-1}$, with the same scaling as $\Phi$. Assuming cosmological parameters as determined by Ref.~\cite{wmap5}, the power in the vector modes is well approximated by~\cite{Lu:2008ju}
 \be
{\cal P}_V\approx 6.5\times10^{-5} {\cal P}_\Phi ~~\mbox{for} ~~ k
\gtrsim k_\mathrm{silk} \approx 0.09\,\text{Mpc}^{-1} \,,
 \ee
so that the amplitude of the metric vector perturbations is nearly 1\% that of the metric scalar modes on small scales.  In general, for a model without baryons,  ${\cal P}_V\approx z_{\rm eq}^{-1}(5.49\, \Omega_{\rm m} h^2-0.13)^{2.33} {\cal P}_\Phi\sim (\ln k)^2/k^4$ for $k\gtrsim k_\mathrm{silk} \approx 0.09 \, \text{Mpc}^{-1}$. On large scales, ${\cal P}_V$ scales like $k$, with a peak in the spectrum around the equality scale. 

Note that these vector degrees of freedom are not associated with the vorticity of the fluid and have no Newtonian counterpart as they induce a non-zero magnetic Weyl curvature. The small-scale behaviour of the second order vector modes can be estimated by replacing the linear transfer function with that given by Halofit~\eqref{halo}, which is depicted on Fig.~\ref{fig-ps}. This gives a more realistic estimation of the relativistic vector modes on small scales.

\subsection{Second order tensor contribution}\label{sec2-3}

The second order tensor modes evolve according to
\bea
 h_{ij}'' + 2{\cal H} h_{ij}' -\Delta h_{ij}  =\Pi_{ij}
\eea 
where the effective anisotropic stress arises from the contribution of non-linear scalar modes and is explicitely given by
\bea
\Pi_{ij}&\equiv& \Big\{-16\Phi\partial_i\partial_j\Phi-8\partial_i\Phi\partial_j\Phi\\ 
&&+ 
\frac{4}{\mathcal{H}^2\Omega_{\rm m}}
\left[ {\cal H}^{2}\partial_i\Phi\partial_j\Phi +2\mathcal{H}\partial_i\Phi\partial_j\Phi' +
\partial_i\Phi'\partial_j\Phi' \right]\Big\}^{TT}\nonumber
\eea
where $TT$ denotes a tensor projection~\cite{Ananda:2006af}.

 In Fourier space, $h_{ij}$ has 2 independent degrees of freedom that can be decomposed as $+,\times$ polarisations as
\begin{equation}\label{eFtens}
h_{ij}(\bm x,\eta) = \int\frac{\dd^3\bk}{(2\pi)^{3/2}}\sum_{\lambda=+,\times}
h_\lambda(\bk,\eta) \varepsilon_{ij}^\lambda(\bk)\, \hbox{e}^{{\rm i}\bk\cdot\bx}\,,
\end{equation}
where $\varepsilon_{ij}^\lambda$ is the polarisation tensor, satisfying $\varepsilon_{ij}^\lambda\delta^{ij}=\varepsilon_{ij}^\lambda k^i=0$ and $\varepsilon_{ij}^\lambda \varepsilon^{ij}_{\lambda'}=\delta^\lambda_{\lambda'}$.

Again, power in each polarization states are identical, thanks to spatial isotropy, and are well approximated by~\cite{Baumann:2007zm,Mollerach:2003nq,Sarkar:2008ii}
\be\label{Ph}
\mathcal{P}_h(k,\eta)=\frac{6C\Delta_\mathcal{R}^4g_\infty}{25}\frac{k_*\left[1-3\frac{j_1(k\eta)}{k\eta}\right]}{k\left[1+7\frac{k_*}{k}+5\left(\frac{k_*}{k}\right)^2\right]^3}\,,
\ee
where $C\sim0.06$ for a scale-invariant spectrum. $j_1$ stands for the $\ell=1$ spherical Bessel function and $k_*=\Omega_{\rm m}h^2$Mpc${}^{-1}$.

The second order gravitational wave background also peaks in power around the equality scale, and is surprisingly larger than its primordial background on these scales. The formula presented in Eq.~\eqref{Ph}, from Ref.~\cite{Mollerach:2003nq}, predicts an excess in power on small scales compared to the more accurate formula of Ref.~\cite{Baumann:2007zm}, but is sufficiently accurate for our purposes (see Ref.~\cite{Sarkar:2008ii} for a direct comparison). 

\subsection{Summary}\label{sec2-4}

\begin{figure}[h]
\begin{center}
\includegraphics[width=\columnwidth]{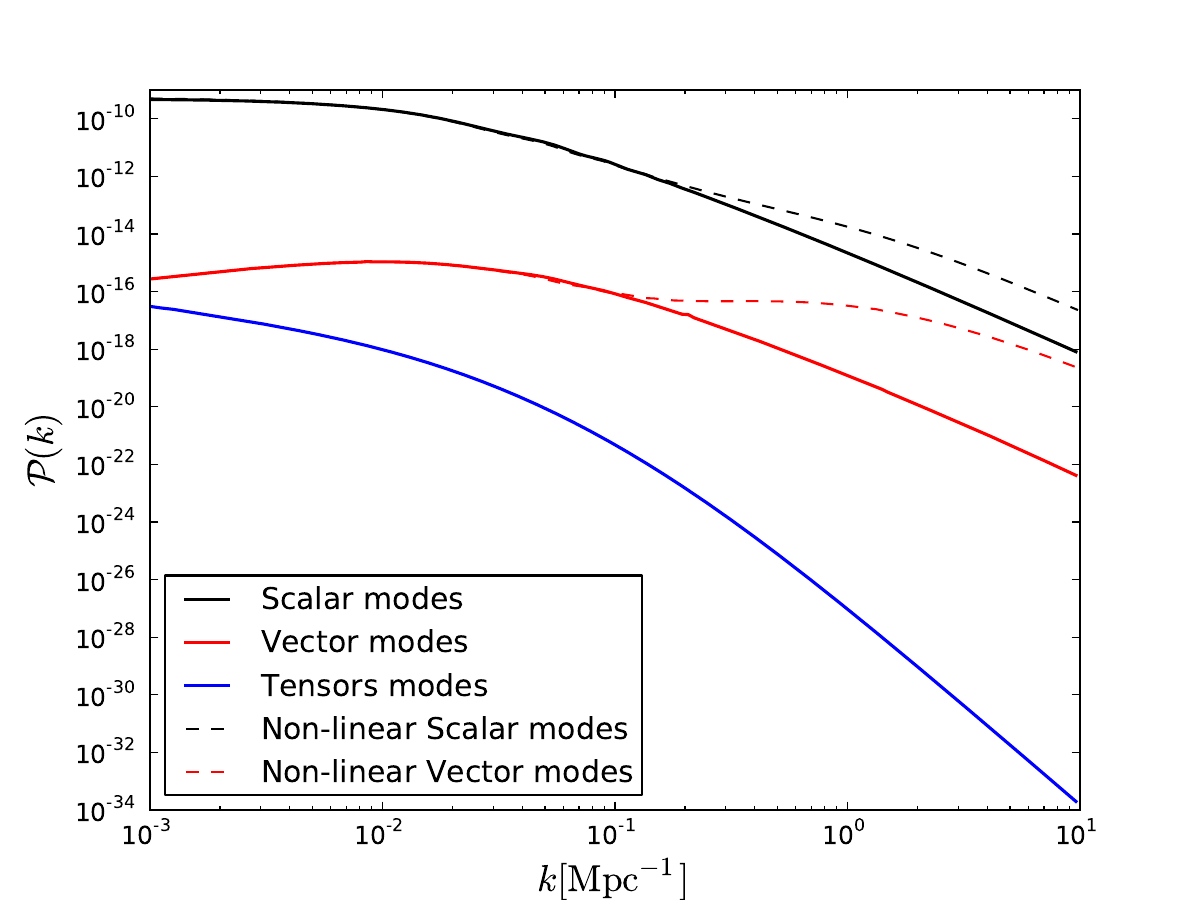}  
\caption{The power spectra of scalar (black line), vector (red line) and tensor (blue line) as a function of the comoving wavenumber $k$. Solid curves  correspond to spectra computed assuming linear scalar modes at first-order, and the dotted curves include power from small-scale clustering estimated from Halofit. } 
 \label{fig-ps}
  \end{center}
\end{figure}

The previous paragraphs give the expession of the power spectra of the scalar modes (both linear and second order), vector modes and tensor modes. Fig.~\ref{fig-ps} depicts these different contributions assuming a flat $\Lambda$CDM background universe with  $\Omega_{\rm m} h^2 = 0.1326$, $\Omega_{\rm b} h^2 = 2.263\times10^{-2}$ and $h = 0.719$ as derived from the WMAP5 best fit model~\cite{wmap5}. We also use the transfer function derived in Ref.~\cite{Eisenstein:1997ik}.

Note that since the amplitudes of vector and tensor modes are small on Mpc scales we do not take into account their non-linear contribution. 

\section{Weak lensing convergence and power spectra}\label{sec3}

\subsection{Generalities}\label{sec3-1}

In the standard lore, the dominant contribution to weak lensing comes from the deflecting potential $\phi$ along a line of sight in the direction 
$n^i$ (see {\em e.g.}, Refs.~\cite{ruth,ub00,ehlers,bartelman}),
\begin{equation}\label{j1}
 \phi = \Phi+\Psi + V_in^i + h_{ij}n^in^j\ ,
\end{equation}
which can be decomposed in contributions arising from the scalar-vector-tensor perturbations of the metric as
\be\label{j2}
\phi=\phi_{_S}+\phi_{_V}+\phi_{_T},
\ee
with $\phi_{_S}=\Phi+\Psi$, $\phi_{_V}=V_in^i $ and $\phi_{_T}=h_{ij}n^in^j$.

The distortion of the shape of background galaxies is described by the Sachs equation~\cite{pubook,ehlers,bartelman} in terms of a Jacobi matrix that can be rescaled, as long as the background spacetime is spatially homogeneous and isotropic~\cite{ppu}, to define the amplification matrix $\mathcal{A}_{ab}$, where the indices refer to the angle coordinates of a unit 2-sphere. At lowest order, it is given by~\cite{pubook,ehlers,bartelman,ppu}
 \begin{equation}
 \mathcal{A}_{ab}= \delta_{ab} - \nabla_a\nabla_b\psi\ ,
\end{equation}
where the lensing potential $\psi$ is obtained by integrating the deflecting potential on the line of sight as
\begin{equation}\label{j3.4}
 \psi(n^i,\chi)
 = \int_0^\chi \frac{f_K(\chi-\chi')}{f_K(\chi)f_K(\chi')}\,
 \phi[f_K(\chi')n^i,\chi']\,
 \dd\chi'\,.
\end{equation}
$\chi$ is the radial coordinate and $f_K$ is defined by
\begin{equation}
 \dd s^2_{(3)} = \dd\chi^2 + f_K^2(\chi)\dd\Omega^2 \ ,
\end{equation}
so that $f_K(\chi)=\chi$ for a spatially Euclidean universe.

The amplification matrix can be decomposed in term of a convergence $\kappa$ and a shear $(\gamma_1,\gamma_2)$ as
\begin{equation}
 \mathcal{A}_{ab} =
 \left(
 \begin{array}{cc}
  1-\kappa- \gamma_1 & -\gamma_2 \\
  -\gamma_2 & 1-\kappa+\gamma_1
 \end{array}
 \right) \ ,
\end{equation}
from which we deduce that
\begin{equation}\label{ch1}
 \kappa(n^i,\chi) = \frac{1}{2}\nabla^2_\perp \psi(n^i,\chi)\,,
\end{equation}
where $\nabla^2_\perp$ is the 2-dimensional Laplacian on the unit 2-sphere. 

The previous expression~(\ref{ch1}) gives the convergence for a single source located at a radial distance $\chi$, or similarly at a redshift $z$. However, observations usually deal with the convergence averaged over a source distribution $n_s$,
\be
 \kappa(n^i) = \int_0^\infty n_s(\chi)\kappa(n^i,\chi)\dd\chi \,,
\ee
where the upper limit of infinity is taken to mean well beyond the source distribution, or the horizon scale. Note that such an averaging over the source distribution is not mandatory if one has distance information about each bin of sources. Using the fact that $\int_0^\infty\dd\chi\int_0^\chi\dd\chi'$ is equivalent to integrate as $\int_0^\infty\dd\chi'\int_{\chi'}^\infty\dd\chi$ we obtain, after exchanging $\chi$ and $\chi'$, the expression
\be\label{eq39}
 \kappa(n^i) = \frac{1}{2}\nabla_\perp^2 \int_0^\infty
 \hat g(\chi)\phi[f_K(\chi')n^i,\chi']\dd\chi
\ee
with
\begin{equation}\label{ch81}
 \hat g(\chi) = \frac{1}{f_K(\chi)}\int_\chi^\infty
 n_s(\chi')\frac{f_K(\chi'-\chi)}{f_K(\chi')}\dd\chi'\ .
\end{equation}
From this, we may also introduce the lensing potential averaged over sources as
\be\label{j3.11}
 \psi(n^i) =  \int_0^\infty
 \hat g(\chi)\phi[f_K(\chi')n^i,\chi']\dd\chi
\ee
in terms of which Eq.~(\ref{eq39}) takes the form 
\be\label{kappa}
\kappa(n^i) = \frac{1}{2}\nabla^2_\perp \psi(n^i)\,.
\ee

The geodesic bundle propagates in the perturbed spacetime, which induces a correction of the redshift of the source, compared to the background redshift. Correcting the redshift in turn corrects the distance to the source, and so adds to the convergence. This affects only the convergence but not the shear (at linear order). Taking into account this effect induces three extra terms at first-order for the convergence: the Sachs-Wolfe and Integrated Sachs-Wolfe terms and a Doppler lensing term  (Refs.~\cite{Bonvin:2008ni,Bonvin:2005,Bacon:2014uja}). The SW and ISW contributions are
\bea
\kappa_{\rm sw}(n^{i},\chi)&=&\left(2-\frac{1}{\mathcal{H}\chi}\right)\Phi(n^{i},\chi), \label{ksw} \\
\kappa_{\rm isw}(n^{i},\chi)&=&2\left(1-\frac{1}{ \mathcal{H}\chi}\right)\int_0^{\chi} d\chi'\,\Phi'(n^{i},\chi') \nonumber\\
&&-\frac{2}{\chi}\int_0^{\chi} d\chi'\,\Phi(n^{i},\chi')\label{kisw}.
\eea
The Doppler contribution, in a spatially Euclidean background, is 
\begin{equation}\label{e1}
\kappa_{v}(n^i,\chi) = -\left[1-\frac{1}{\chi \mathcal{H}(\chi)}\right]n^i v_{i}\, ,
\end{equation}  
for $\bm n$ pointing in the direction of observation, and
with $v_i$ given by Eq.~(\ref{j2.2}). This contribution to the convergence was first identified in~\cite{Bonvin:2008ni,Bonvin:2005}, and investigated in more detail in~\cite{Bacon:2014uja,voidslensing}. Note that when using these formula, the comoving distance to a source $\chi$ should be calculated from the background distance-redshift relation using the observed redshift (and not the unphysical background redshift).

\subsection{Different contributions to the convergence}\label{sec3-2}

As discussed in \S~\ref{sec2}, we have 3 contributions to the convergence that arise from the scalar, vector and tensor contributions to Eqs.~(\ref{j1}-\ref{j2}), to which we need to add the two Sachs-Wolfe terms and an important first-order contribution induced by the Doppler effect~\cite{Bonvin:2008ni}. 

It follows that the observed weak lensing convergence has 4 contributions given by:

\noindent\emph{at first-order}
\begin{eqnarray}
 \kappa_{_S}(n^i) &=& \frac{1}{2}\nabla^2_\perp \int_0^\infty\dd\chi\,
 \hat g(\chi)\left[\Phi(n^i,\chi)+\Psi(n^i,\chi)\right]\ \ \ \ \ , \\
 \kappa_v(n^i) &=& \int_0^\infty\dd\chi\, n_s(\chi) \left[\frac{1}{\chi \mathcal{H}(\chi)}-1\right]n_i v^{i}(n^i,\chi)\label{corrdop}\\
 \kappa_{\rm sw}(n^{i})&=&\int_0^{\infty}d\chi n_{s}(\chi)\left(2-\frac{1}{\mathcal{H}\chi}\right)\Phi(n^{i},\chi) \label{ksw} \\
\kappa_{\rm isw}(n^{i})&=&2\int_0^{\infty} d\chi\ \hat g_{\rm isw1}(\chi)\Phi'(n^{i},\chi) \label{corrisw}\\ \nonumber
&&-2\int_0^{\infty} d\chi \hat g_{\rm isw2}(\chi)\Phi(n^{i},\chi)\label{kisw},\label{jeqv} 
 \eea
 where $$\hat g_{\rm isw1} = \left(1-\frac{1}{ \mathcal{H}\chi}\right)\int_{\chi}^{\infty}d\chi' n_{s}(\chi')$$ $$\hat g_{\rm isw2}  = \frac{1}{\chi}\int_{\chi}^{\infty}d\chi'n_{s}(\chi')$$ 
 and \emph{at second-order}
 \bea
 \kappa_{_V}(n^i) &=& \frac{1}{2}\nabla^2_\perp \int_0^\infty\dd\chi\,
 \hat g(\chi)n_iV^i(n^i,\chi)\,,\\
 \kappa_{_T}(n^i) &=& \frac{1}{2}\nabla^2_\perp \int_0^\infty\dd\chi\,
 \hat g(\chi)n_in_jh^{ij}(n^i,\chi) \,.
\end{eqnarray}
At second-order in vector and tensor modes, there are also the counterparts of the correction terms given in Eqs.~(\ref{corrdop}),(\ref{ksw}),(\ref{corrisw}) (see Refs.~\cite{Obinna:2012,Obinna:2014}) but we are not considering them in this analysis. As already mentioned, we are only taking into account the dynamically induced vector modes.
 Note also that in these expressions, the variables are evaluated along the light cone and considered as function of the radial distance $\chi$ and the angular position $n^i$ only. Given a source distribution, the left-hand side are purely function of position on the sky.

\begin{figure*}
  \includegraphics[width=1\columnwidth]{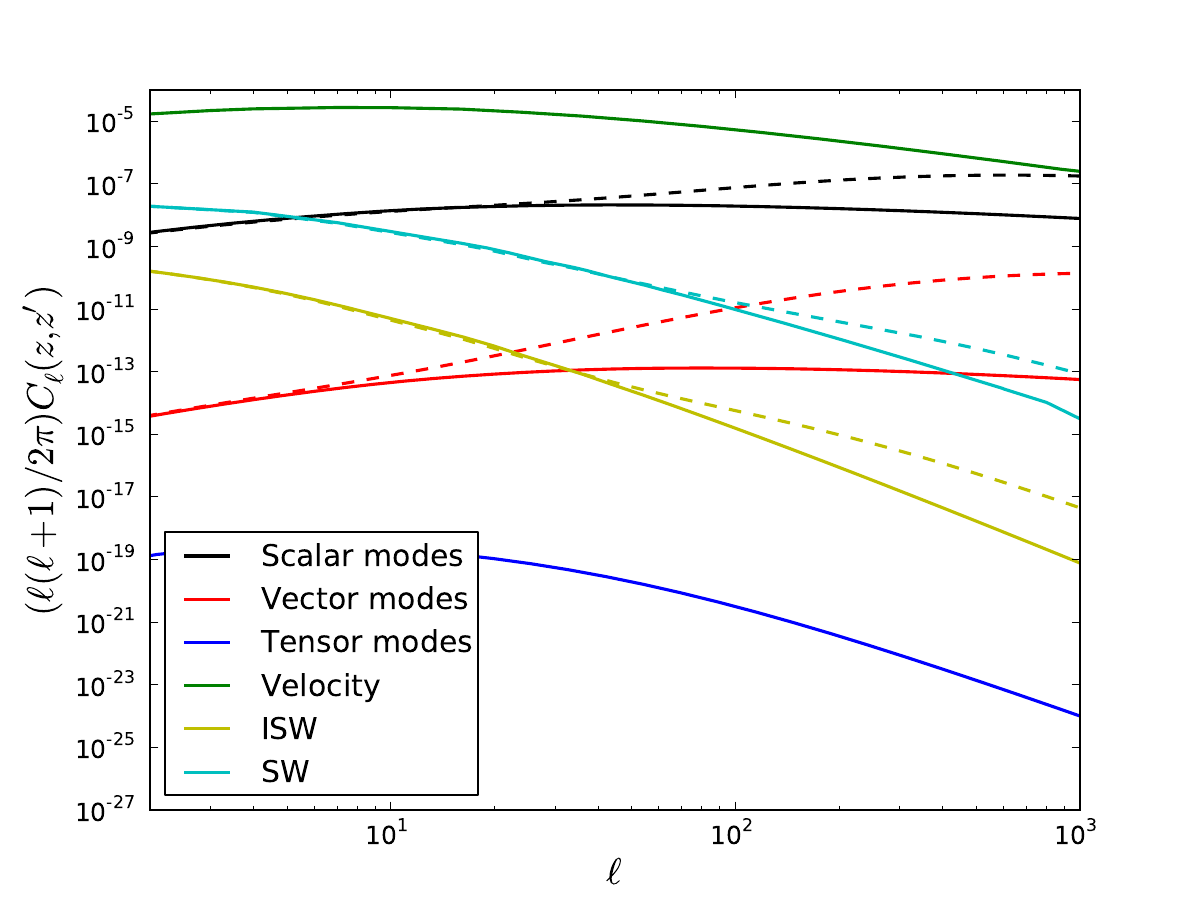}  
  \includegraphics[width=1\columnwidth]{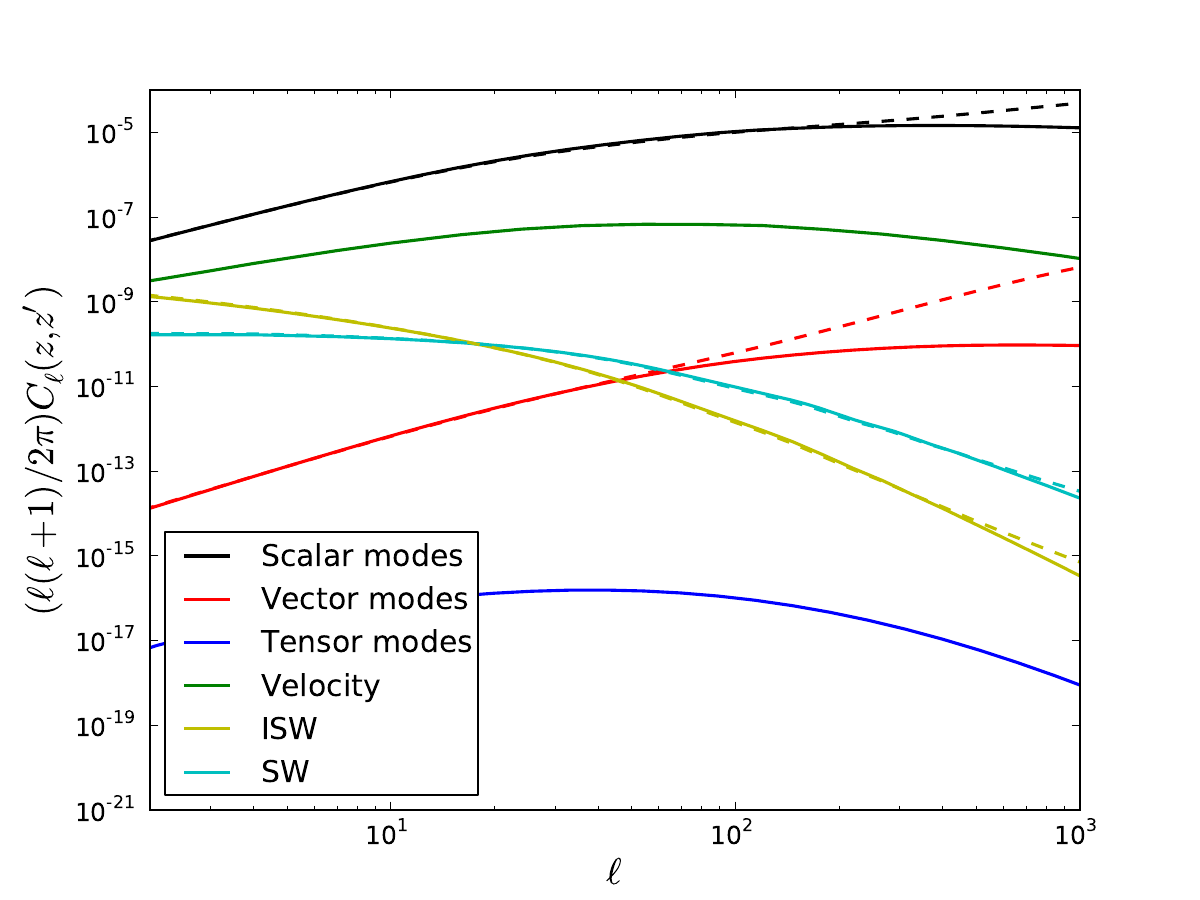}  
  \caption{Lensing angular power spectra of the density contrast (scalars - black line), the Doppler contribution (green), vectors (red line), tensors (blue line), ISW term (yellow line) and SW term (cyan line), $C^{\kappa \kappa}_{\ell}(z,z')$, at $z=z' = 0.1$ (left) and $z=z'=1.0$ (right). The dashed lines represent the non-linear evolution estimated using Halofit. $z$ and $z'$ are the redshifts of the sources on the two lines of sight.} 
  \label{p12} 
\end{figure*}

\subsection{Expression of the power spectra}\label{sec3-3}

Given the previous expressions, one can deduce the angular power spectra of these different contributions to the convergence. To that purpose, we decompose each variable in spherical harmonics. For each contribution, the deflecting potential~(\ref{j3.4}) can be expanded as
\be\label{eq:sh}
\psi(\bm n;\chi)=\sum_{\ell m}\psi_{\ell m}(\chi)Y_{\ell m}(\bm n)\,,
\ee
where $\bm n$ is the position on the celestial 2-sphere, for a source located at $\chi$. Taking into account spatial isotropy, its angular power spectrum  is defined as
\be
\langle \psi_{\ell m}(\chi)\psi^*_{\ell' m'}(\chi')\rangle=C_\ell^{\psi\psi}(\chi,\chi')\delta_{\ell\ell'}\delta_{mm'}\,.
\ee
Given Eq.~(\ref{kappa}), the coefficients of the expansion of the shear are related to the $\psi_{\ell m}$ by 
\be
\kappa_{\ell m}=-\sfrac{1}{2}\ell(\ell+1)\psi_{\ell m}\, ,
\ee
which implies that the angular power spectra of the cosmic convergence and deflecting potential are related by
\begin{equation}
 C_{\ell}^{\kappa\kappa}=\sfrac{1}{4}{\ell^2(\ell+1)^2}C_{\ell}^{\psi\psi}\,.
\end{equation}
The power spectra are related to the real space angular correlation function,
\be
C^{\psi\psi}(\bm n\cdot\bm n';\chi,\chi')=\langle\psi(\bm n,\chi)\psi(\bm n',\chi')\rangle\,
\ee
by
\be
C^{\psi\psi}(\bm n\cdot\bm n';\chi,\chi')=\sum_{\ell=0}^{\infty}\frac{2\ell+1}{4\pi}C_\ell^{\psi\psi}(\chi,\chi')P_\ell(\bm n\cdot\bm n')\, ,
\ee
where $P_\ell$ stands for the Legendre polynomials. 

When the integration over the source distribution is included ({\em i.e.} using the expressions~(\ref{eq39}-\ref{kappa})), one obtains similar expressions for the angular power spectra but with an extra integration over the sources  distribution so that the dependence in $\chi$ disappears.

The derivation of the angular power spectra is detailed in Appendices~\ref{appA}, \ref{appD}, \ref{appE}, \ref{appB} and~\ref{appC} respectively for the velocity term, ISW term, SW term, the vector and tensor modes.

After integrating over the sources distribution, all power spectra (see Eqs.~(\ref{eA4}), (\ref{eB6}), (\ref{eC14}), (\ref{eD15}), (\ref{swD1}), (\ref{iswE1}), (\ref{iswE2}) and~(\ref{iswE3})) can all be written as
\bea\label{jspectre}
C_\ell^{\psi_X\psi_X}&=&\left[A^{(s)}_\ell\right]^2 \int_0^\infty\hat g(\chi)\dd\chi\int_0^\infty\hat g(\chi')\dd\chi'\nonumber \\
&&\int \frac{\dd k}{k}\frac{j_\ell(k\chi)}{(k\chi)^{s}} \frac{j_\ell(k\chi')}{(k\chi')^{s}} {\cal P}_X(k,\chi,\chi')\, ,
\eea
with
\be
A^{(s)}_\ell = \sqrt{\frac{16\pi}{N_s^2 F_{s}}\frac{(\ell+s)!}{(\ell-s)!}}
\ee
where $X=\{S, V, T\}$, corresponding to  $s=0,1,2$. The power spectra of each mode, ${\cal P}_S={\cal P}_\Phi$ etc., are respectively given by Eqs.~(\ref{e.powerS}), (\ref{e.powerV}) and~(\ref{Ph}) and we have replaced $\eta=\eta_0-\chi$ by $\chi$ since this the integral is evaluated on the past lightcone. The numbers $F_s=(1,2,8)$ for $s=(0,1,2)$ and  $N_s=(1,2,2)$ is the number of polarisations of each mode. The Doppler contribution ($X=v$)  takes a similar form (see Appendix~\ref{appA}) with $A^{(s)}_\ell\rightarrow \sqrt{4\pi}A$, $\hat g(\chi)\rightarrow F(\chi)$, $j_\ell(k\chi)/(k\chi)^s\rightarrow j_\ell'(k\chi)$ and ${\cal P}_v\rightarrow k^2{\cal P}_\Phi$. The two contributions from ISW and SW terms are both similar to the scalars modes with $s = 0$, $F_s = 1$ and $N_s = 1$ except that for SW $A^{(s)} = \sqrt{4\pi}$ whereas that of ISW is the same as the scalar modes (see Appendices~\ref{appD} to~\ref{appE}). 

Each spectrum can be written in terms of a transfer function $T_X(k,\eta)$ which is  normalized to unity at early times as
\be
\mathcal{P}_X(k,\eta,\eta')=\mathcal{P}_{X,i}(k)T_X(k,\eta)T_X(k,\eta')\,.
\ee
This implies that Eq.~(\ref{jspectre}) factors as
\bea\label{ch3}
C_\ell^{\psi_X\psi_X}&=& \left[A^{(s)}_\ell\right]^2 \int_0^\infty\frac{\dd k}{k}{\cal P}_{X,i}(k)\nonumber\\
&& \left[\int_0^\infty\dd\chi\,\hat g(\chi)\frac{j_\ell(k\chi)}{(k\chi)^{s}}T_X(k,\chi)\right]^2\,.
\eea
Similarly, the convergence angular power spectra, not integrated over the sources distribution, takes the form 
\bea
C_\ell^{\psi_X\psi_X}(\chi_S,\chi'_S)&&=\left[A^{(s)}_\ell\right]^2
 \int_0^{\chi_S}\dd\chi\frac{f_K(\chi_S-\chi)}{f_K(\chi_S)f_K(\chi)}\nonumber \\
&&\int_0^{\chi_S'}\dd\chi'\frac{f_K(\chi_S'-\chi')}{f_K(\chi_S')f_K(\chi')} \\
&&\int_0^\infty\frac{\dd k}{k} \frac{j_\ell(k\chi)}{(k\chi)^{s}} \frac{j_\ell(k\chi')}{(k\chi')^{s}}{\cal P}_X(k,\chi,\chi').\nonumber
\eea

\begin{figure*}
  \includegraphics[width=0.65\columnwidth]{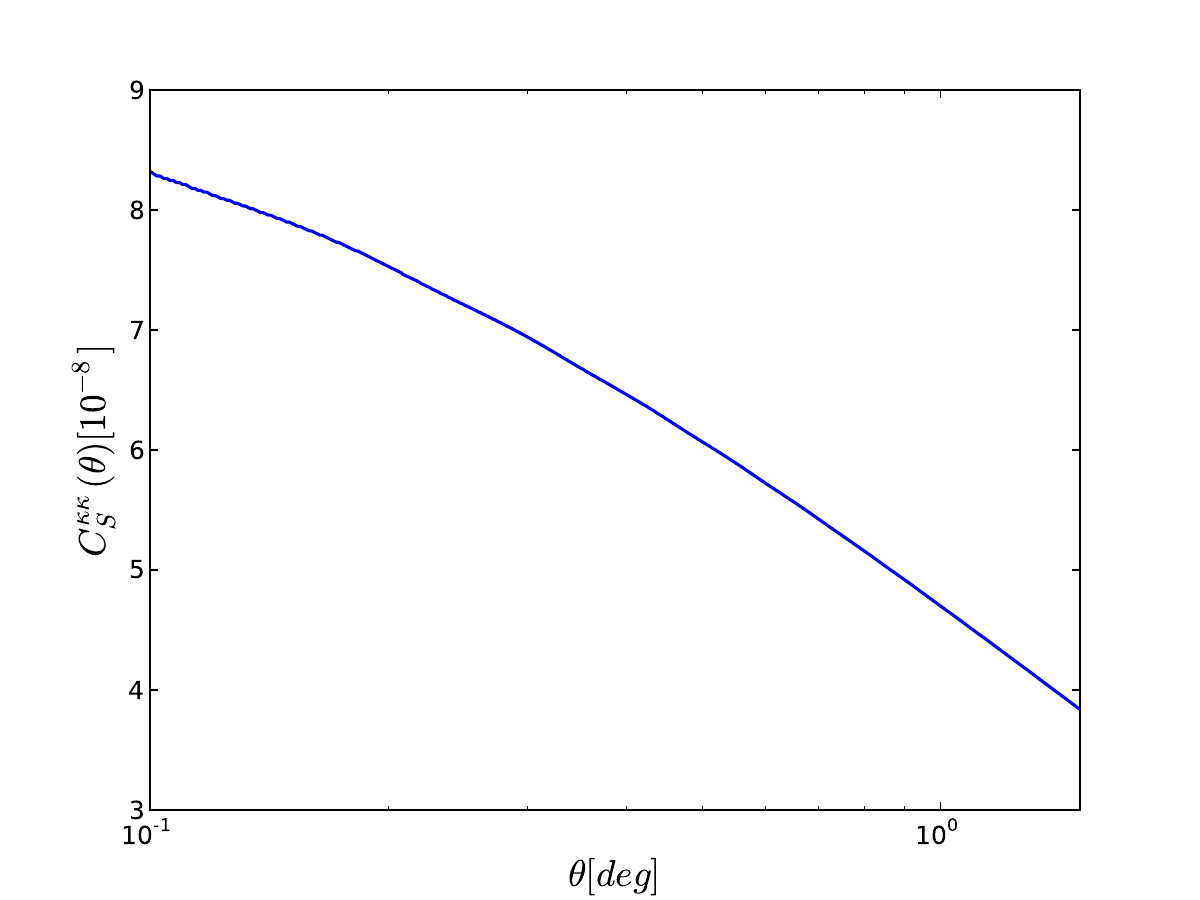} 
         \includegraphics[width=0.65\columnwidth]{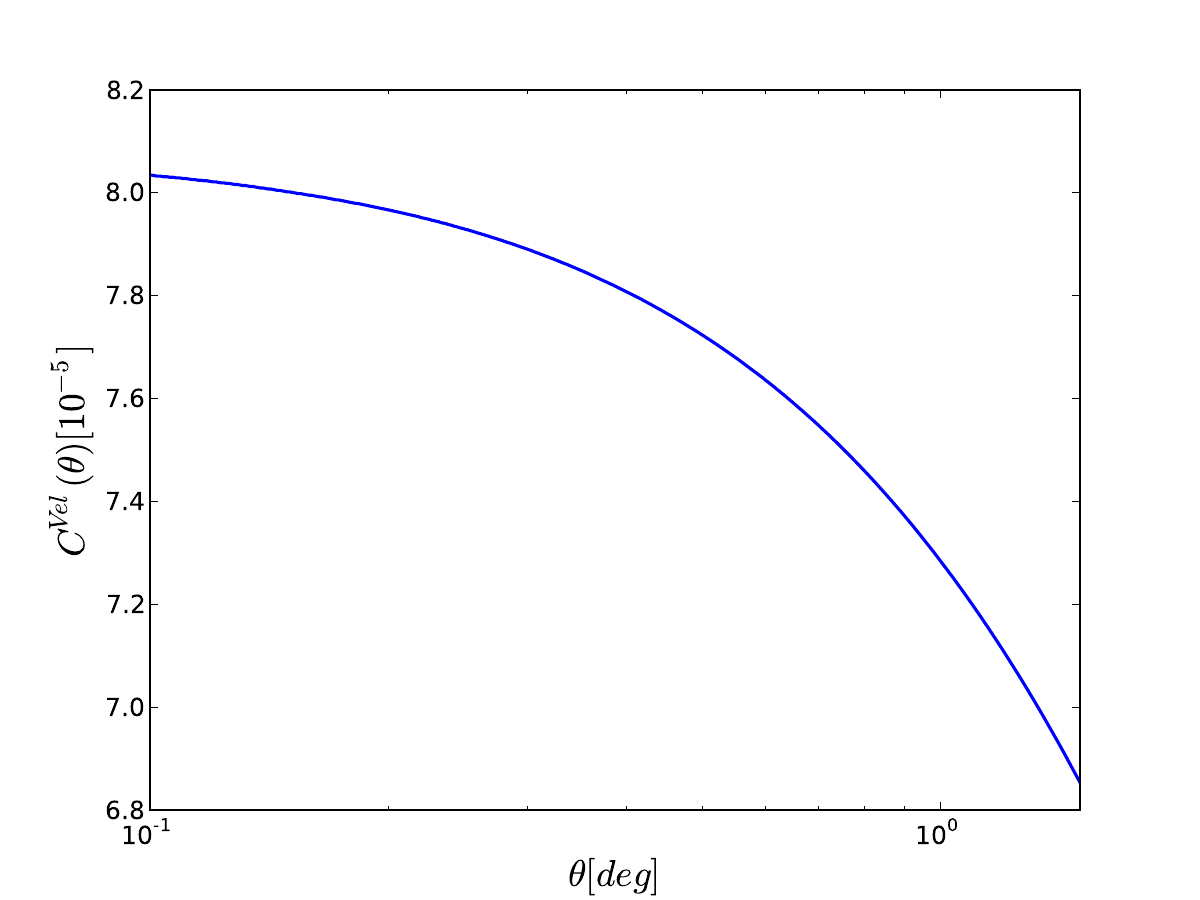}
     \includegraphics[width=0.65\columnwidth]{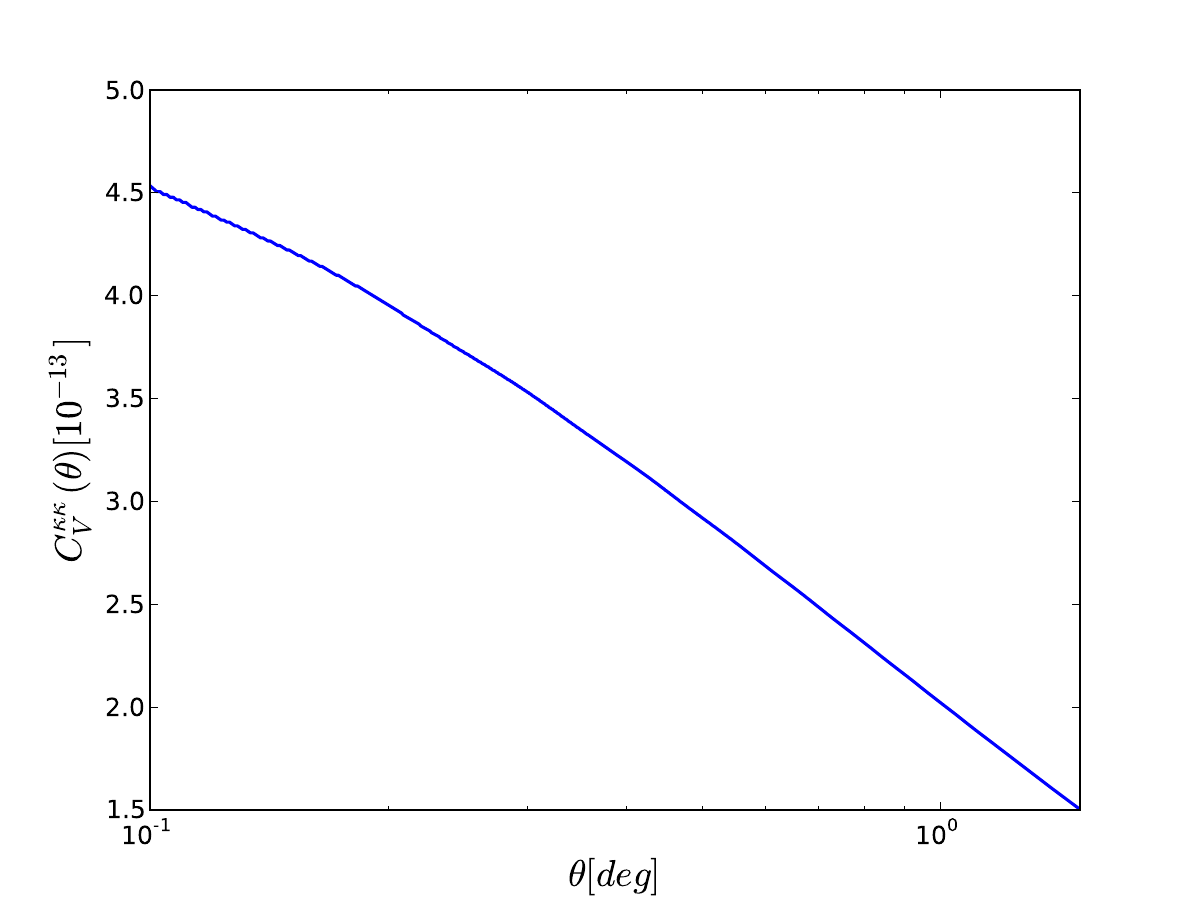} 
  \caption{Real space angular correlation function, $C^{\kappa \kappa}(\theta)$, at $z=z' = 0.1$ for the scalars, Doppler, vectors from left to right. Note that in this regime the Doppler lensing is dominant~\cite{Bonvin:2008ni,Bacon:2014uja}.} 
 \label{p8} 
\end{figure*}

\begin{figure*}
  \includegraphics[width=0.65\columnwidth]{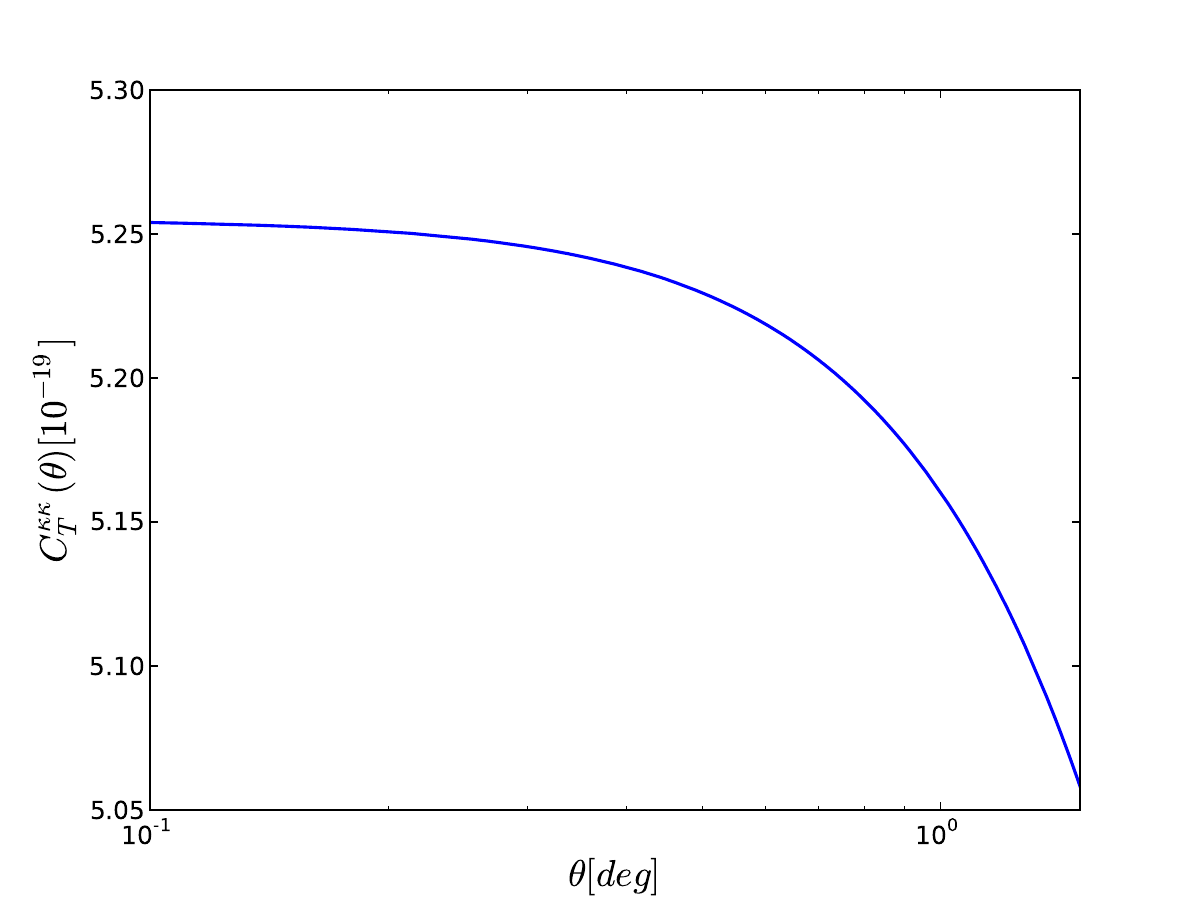} 
         \includegraphics[width=0.65\columnwidth]{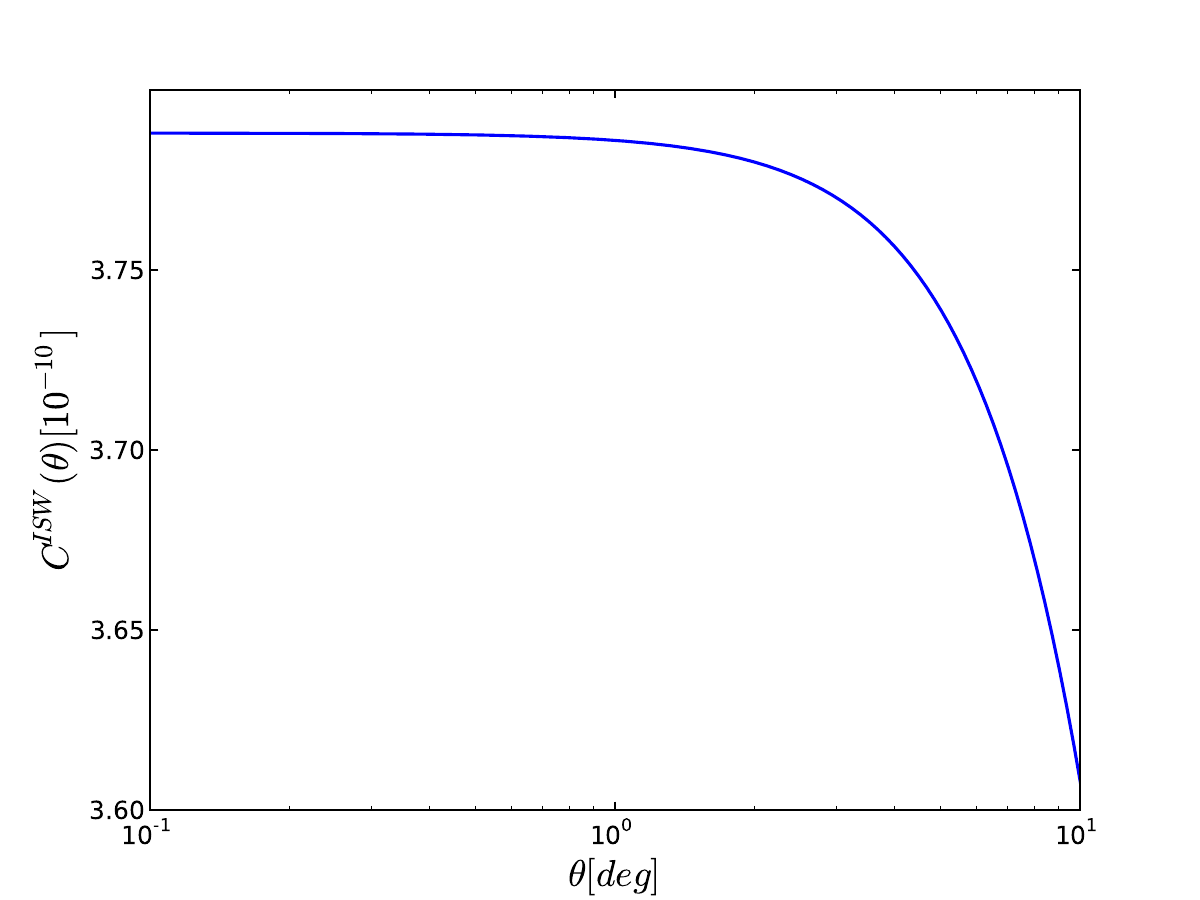}
     \includegraphics[width=0.65\columnwidth]{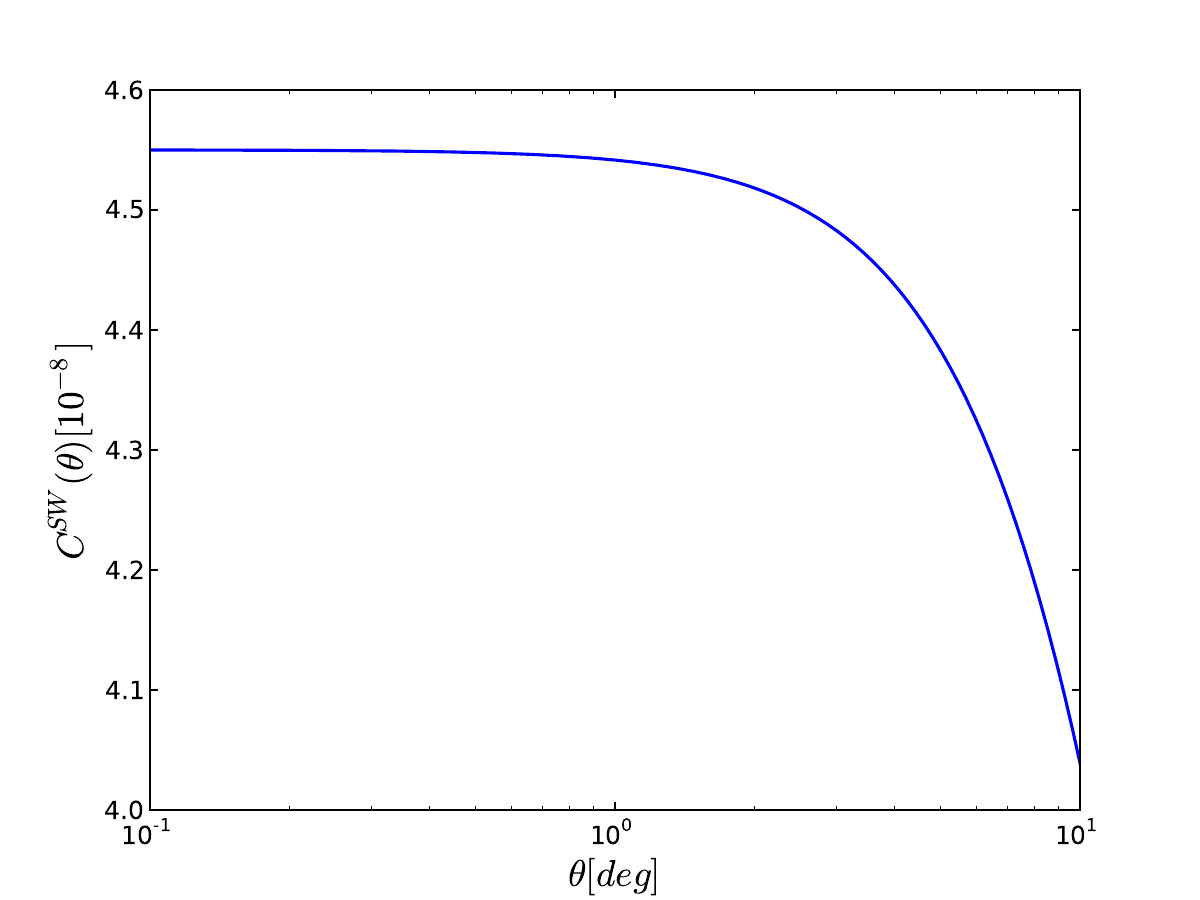} 
  \caption{Real space angular correlation function, $C^{\kappa \kappa}(\theta)$, at $z=z' = 0.1$ for the tensors, ISW and SW from left to right.} 
 \label{p81} 
\end{figure*}

Since integrating the Bessel function in Eq.~(\ref{ch3}) is computationally expensive and since the sources distribution is slowly varying over long distances, we shall resort to a Limber approximation which is a good approximation as at large $\ell$.  In such an approximation,  $k\chi \simeq (\ell + 1/2)$~so that~\cite{LoVerde:2008re,bpu} we have the property
\be
 \frac{2}{\pi}\int_0^\infty{k^2 \dd k}f(k) j_{\ell}(k\chi)j_{\ell}(k\chi') = \frac{\delta(\chi-\chi')}{\chi^2}f[(\ell + 1/2)/\chi]
 \ee
which is accurate to $\mathcal{O}[(\ell + 1/2)^2]$ and is sufficient for our purposes. 
We then find
 \begin{eqnarray}\label{ch9n}\
C_\ell^{\psi_X\psi_X}&=&\frac{64\pi^2}{N_s(2\ell+1)^{3+2s}}\frac{(\ell+s)!}{(\ell-s)!}\times \\
     && \!\!\!\!\!\!\!\!\!\!   \int_{0}^{\infty} \dd \chi\,\chi \mathcal{P}_{X,i}\left[\frac{2\ell+1}{2\chi}\right]\,
   \hat g(\chi)^2 T^2\left[\frac{2\ell+1}{2\chi},\chi\right]\,.\nonumber
\end{eqnarray}

\section{Weak lensing from second-order modes}\label{sec4}

The previous expressions allow us to compute numerically the angular power spectra of the 6 contributions to the cosmic convergence in particular to estimate the typical magnitude of the non-linear terms which we compare to the standard term $\kappa_S$, the Doppler term $\kappa_v$, ISW term $\kappa_{\rm isw}$ and SW term $\kappa_{\rm sw}$, which allows us to discuss whether assuming $\kappa_{\rm observation}=\kappa_S+\kappa_v$ is a good approximation to interpret the weak lensing observations. Since the two point function can be computed in real space ({\em i.e.}, the correlation function $C(\theta,z,z')$) or in harmonic space  ({\em i.e.}, the angular power spectrum $C_\ell(z,z')$), we shall use the two representations.

\begin{figure*}
  \includegraphics[width=.66\columnwidth]{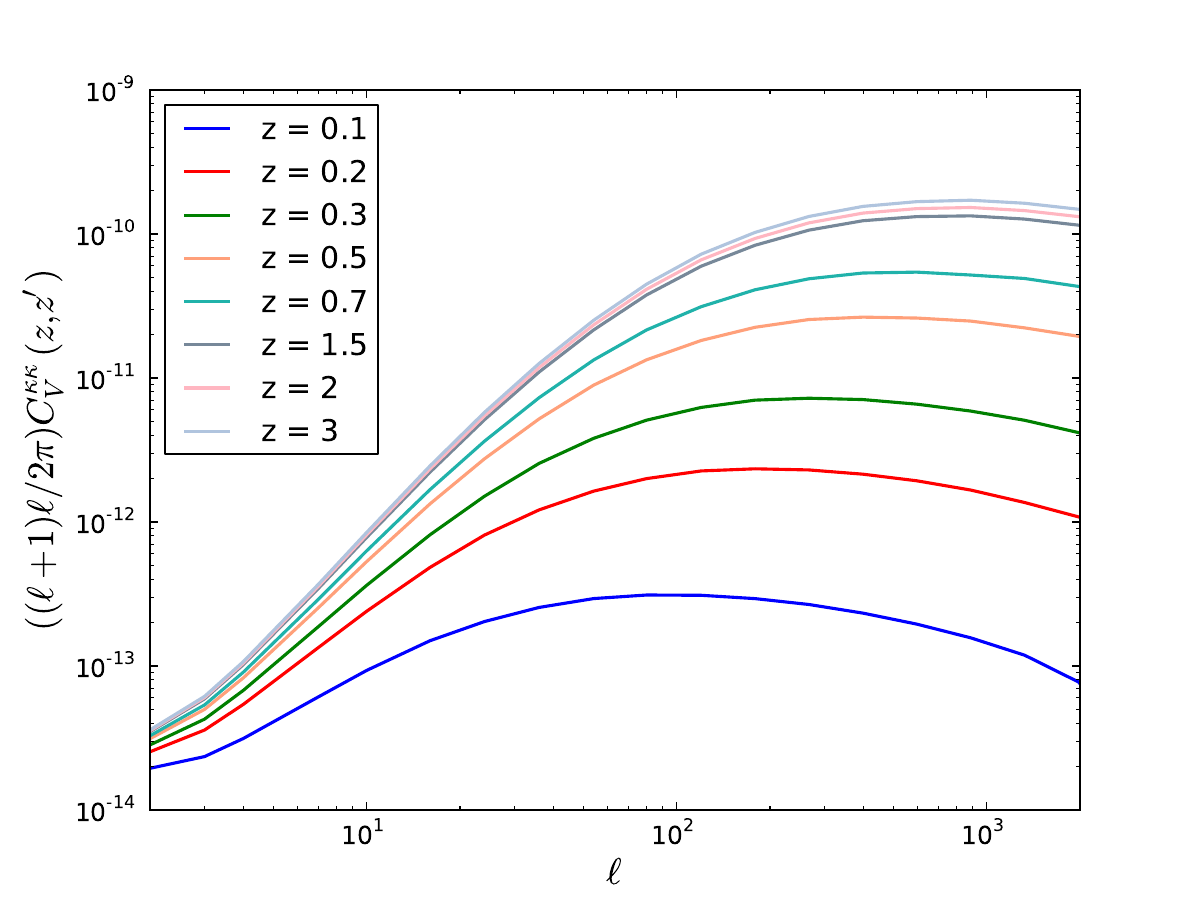}
  \includegraphics[width=.66\columnwidth]{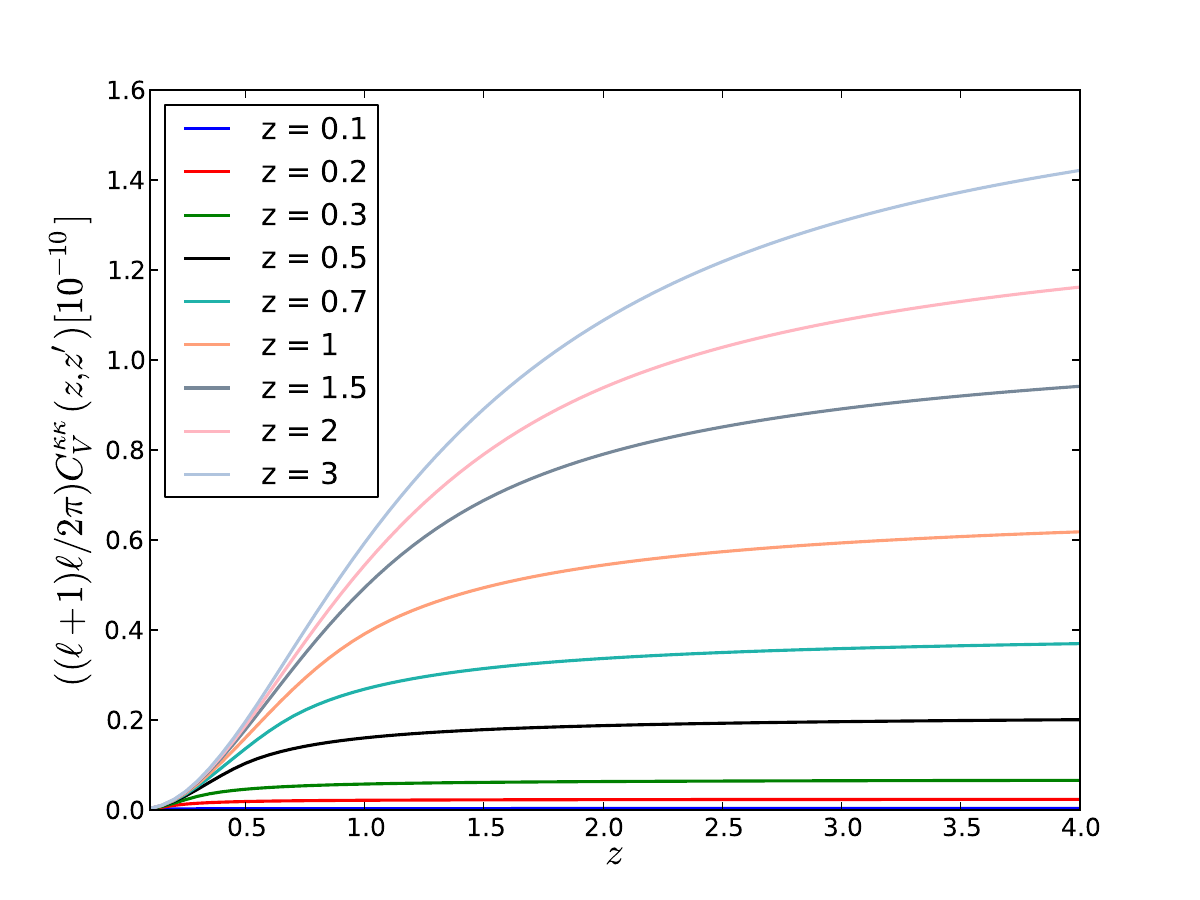} 
  \includegraphics[width=.66\columnwidth]{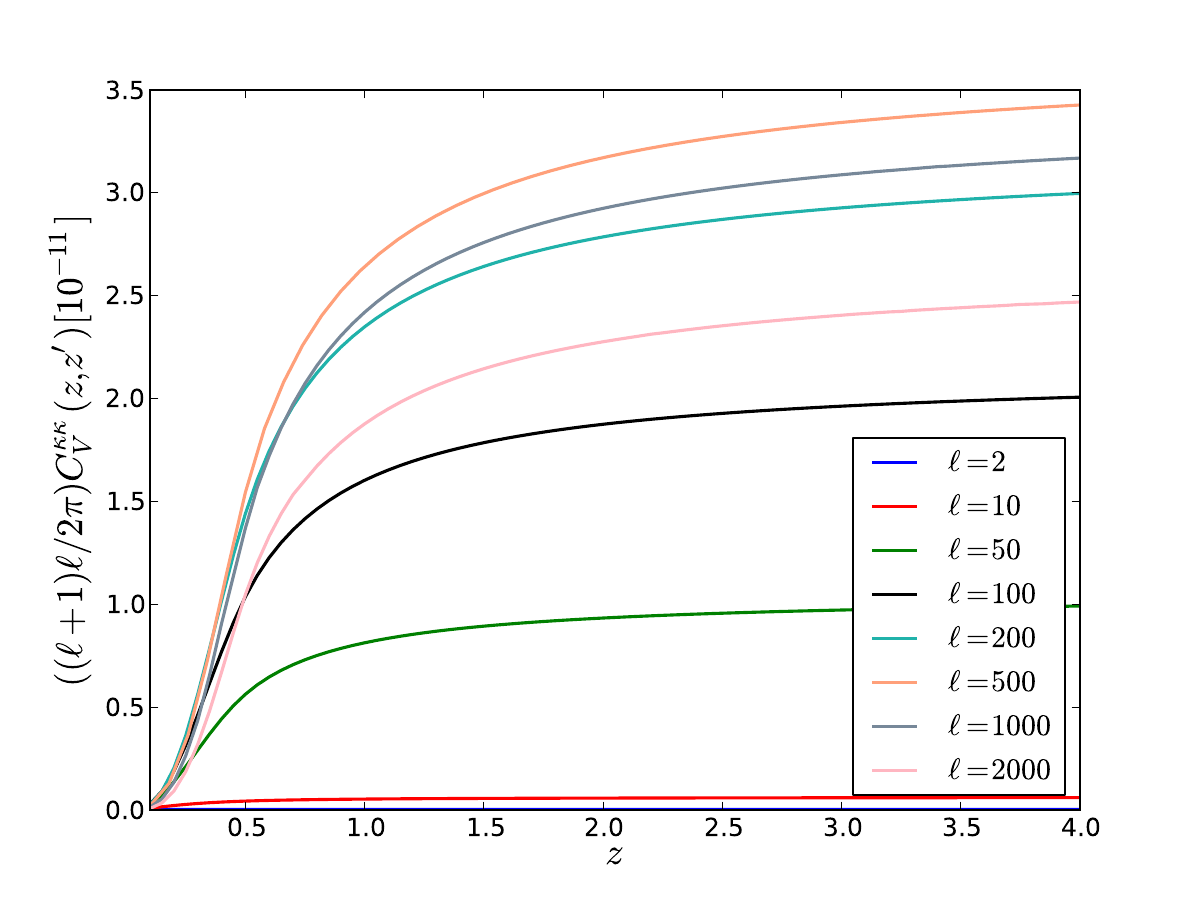}    
  \caption{Angular power spectra of the vectors at different redshifts. (left) as a function of $\ell$ for $z'=1$; (middle) as a function of $z$ for different $z'$ for $\ell = 100$ and (right) as a function of $z$ for different multipole $\ell$ with $z^{\prime} = 0.5$.} 
  \label{p-vec-cl} 
\end{figure*}

\begin{figure}
  \includegraphics[width=\columnwidth]{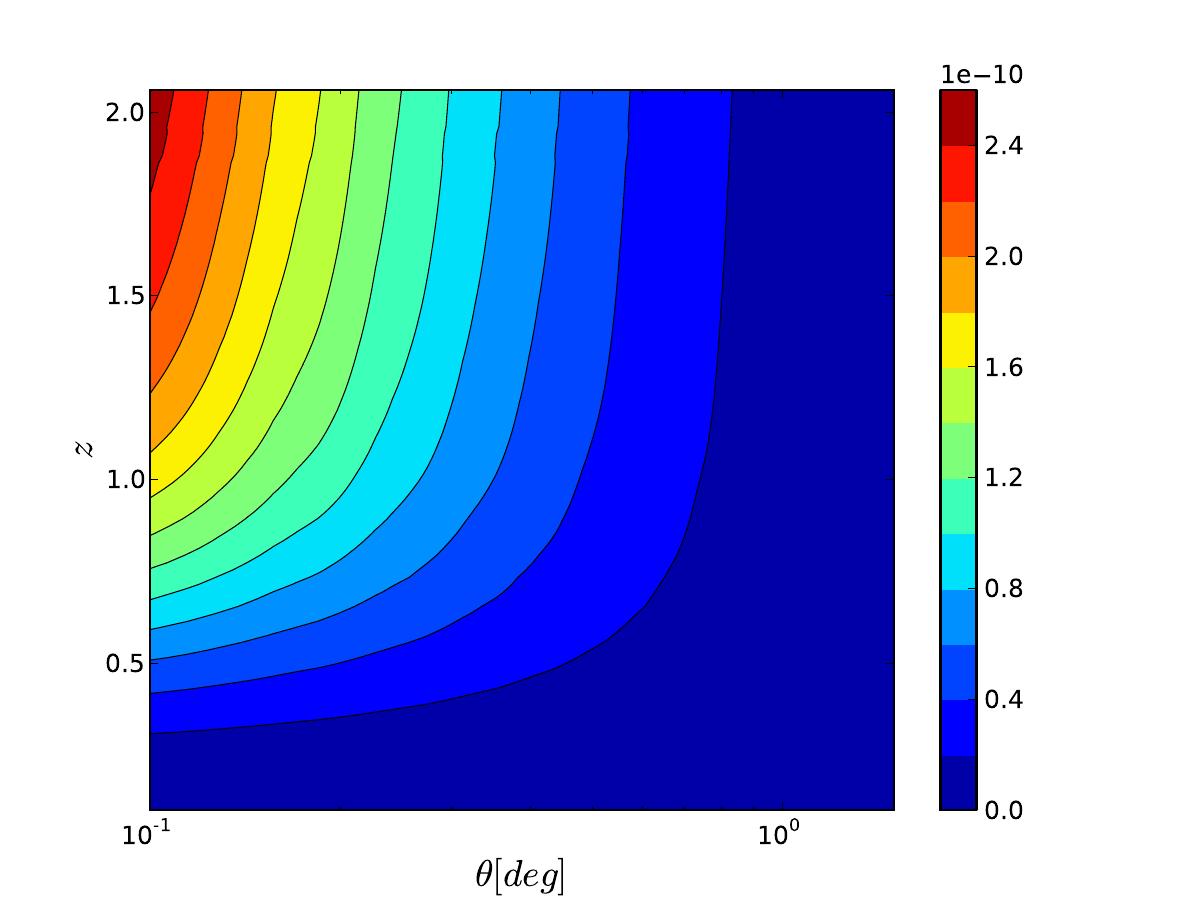}
 \caption{\label{p-vec-cl2} Amplitude of the  angular power spectra of the vectors in real space with $z'=1$. } 
\end{figure}

\subsection{Behaviour of the different contributions}

We start by comparing in Fig.~\ref{p12} the different contributions to the lensing angular power spectra without integrating over the sources distribution and assuming that the sources on the sky are located at the same redshift  in $z=z' = 0.1$ or  $z=z'=1.0$. We recover that the velocity contribution dominates at low redshift~\cite{Bonvin:2008ni} and that the gravity waves contribution is always negligible~\cite{Sarkar:2008ii}. The results shown in Fig.~\ref{p12} also suggest that there is a range in multipoles $\ell$ ($\ell \ge 50 $) where the second order vector modes become more significant than both of the Sachs-Wolfe terms. A similar computation in real space, assuming $z=z'=0.1$ is depicted in Fig.~\ref{p8} and Fig.~\ref{p81}.

\begin{figure*}
\begin{center}
\includegraphics[width=\columnwidth]{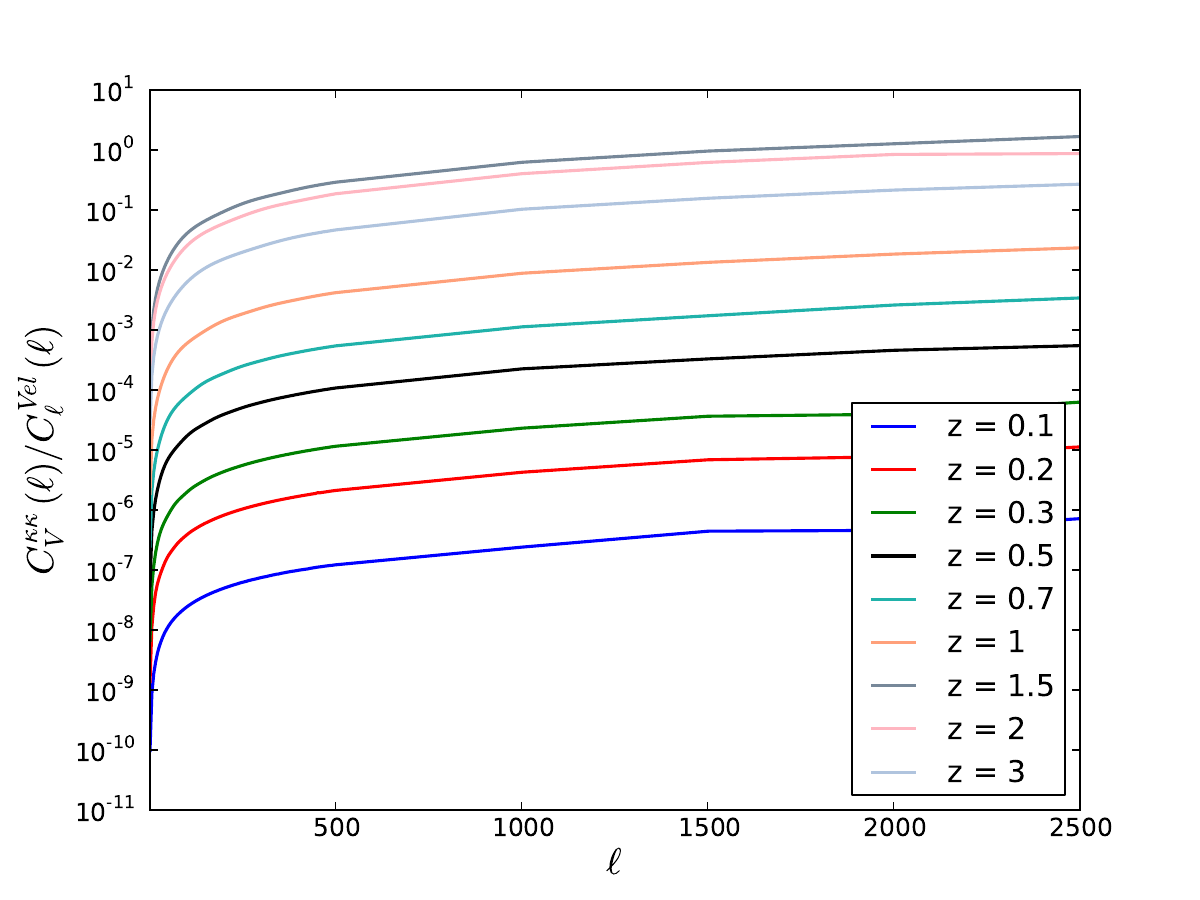}
\includegraphics[width=\columnwidth]{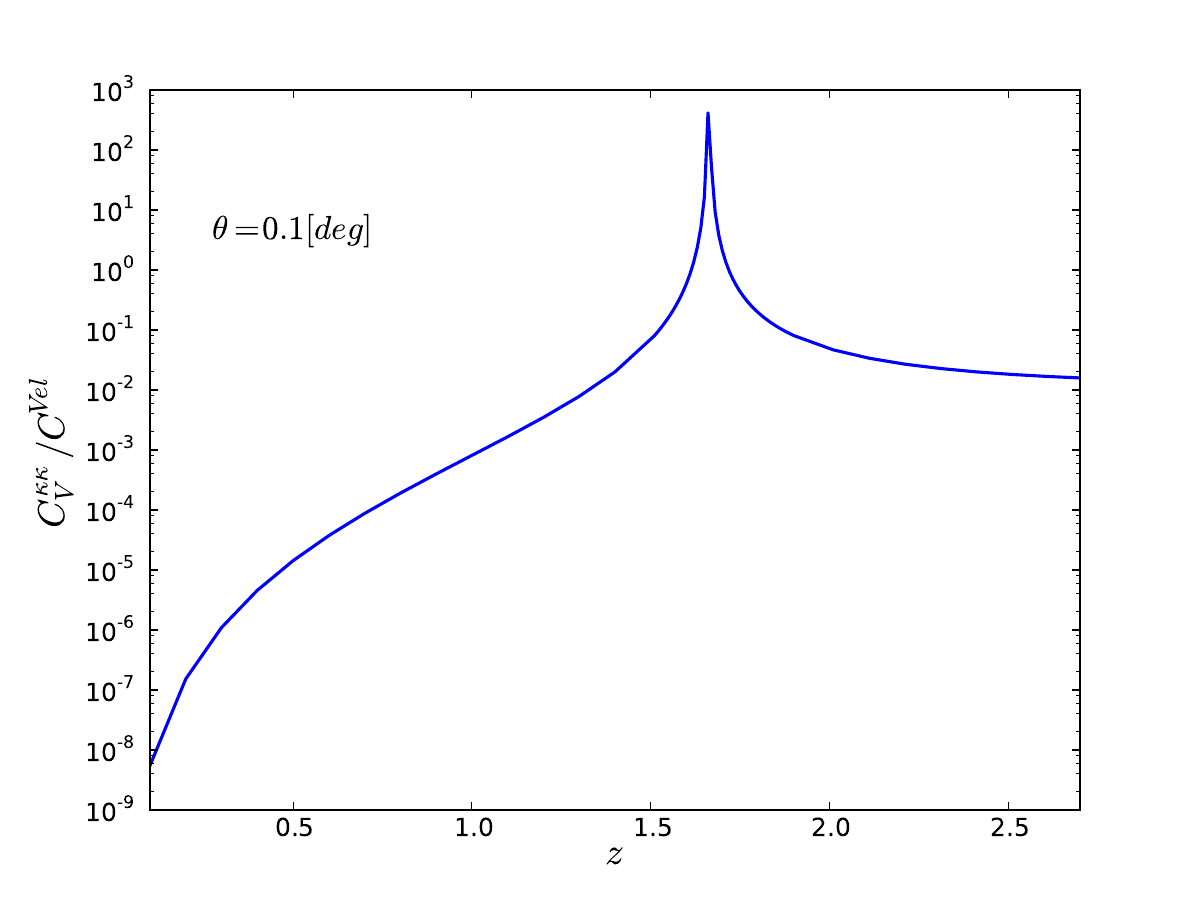}
\caption{(left) : Ratio between the convergence  from the vector mode background to the Doppler convergence  $C^{Vel}_{\ell}$. At moderate redshifts the second-order vectors are larger than the first-order contribution from the Doppler convergence. (right) : Ratio of the two correlation functions (vector modes and the velocity) $C^{\kappa \kappa}_{V}/C^{Vel}$ as a function of $z$ ($z = z^{\prime}$) where $\theta = 0.1$ degree.}
\label{ratio}
\end{center}
\end{figure*}

Focusing on the contribution of the vector modes, Fig.~\ref{p-vec-cl} shows how the amplitude of the angular power spectrum $C_{V,\ell}^{\kappa\kappa}(z,z')$ depends on the redshifts of the background galaxies and on the scale, while Fig.~\ref{p-vec-cl2} shows the similar information in real space, {\em i.e.}, $C_{V}^{\kappa\kappa}(\theta,z,z')$. Fig.~\ref{ratio} shows the ratio of the vectors to the Doppler term, which shows that at intermediate redshifts the second-order frame dragging effects dominate the linear Doppler lensing. It is noted that although, we think that higher order contributions from the vector modes will be subdominant, this issue still needs to be addressed. We have included the non-linear power spectrum using Halofit to compute the vector modes shown in Fig.~\ref{p12}, but for the rest of the calculations for consistency, we only use the linear power spectrum to compute the contribution of the vector modes to the cosmic convergence. 

\subsection{Source distributions}

The source distribution depends on the survey and is described through the function $n_{s,\chi}(\chi)$ or an equivalent function $n_{s,z}(z)$ in redshift space, where $n_{s,\chi}(\chi)\dd\chi=n_{s,z}(z)\dd z$. These distributions are normalised to unity. This then defines the lensing weight function $\hat g$, as shown in Eq.~(\ref{ch81}). 

\begin{figure*}
\begin{center}
\includegraphics[width=.66\columnwidth]{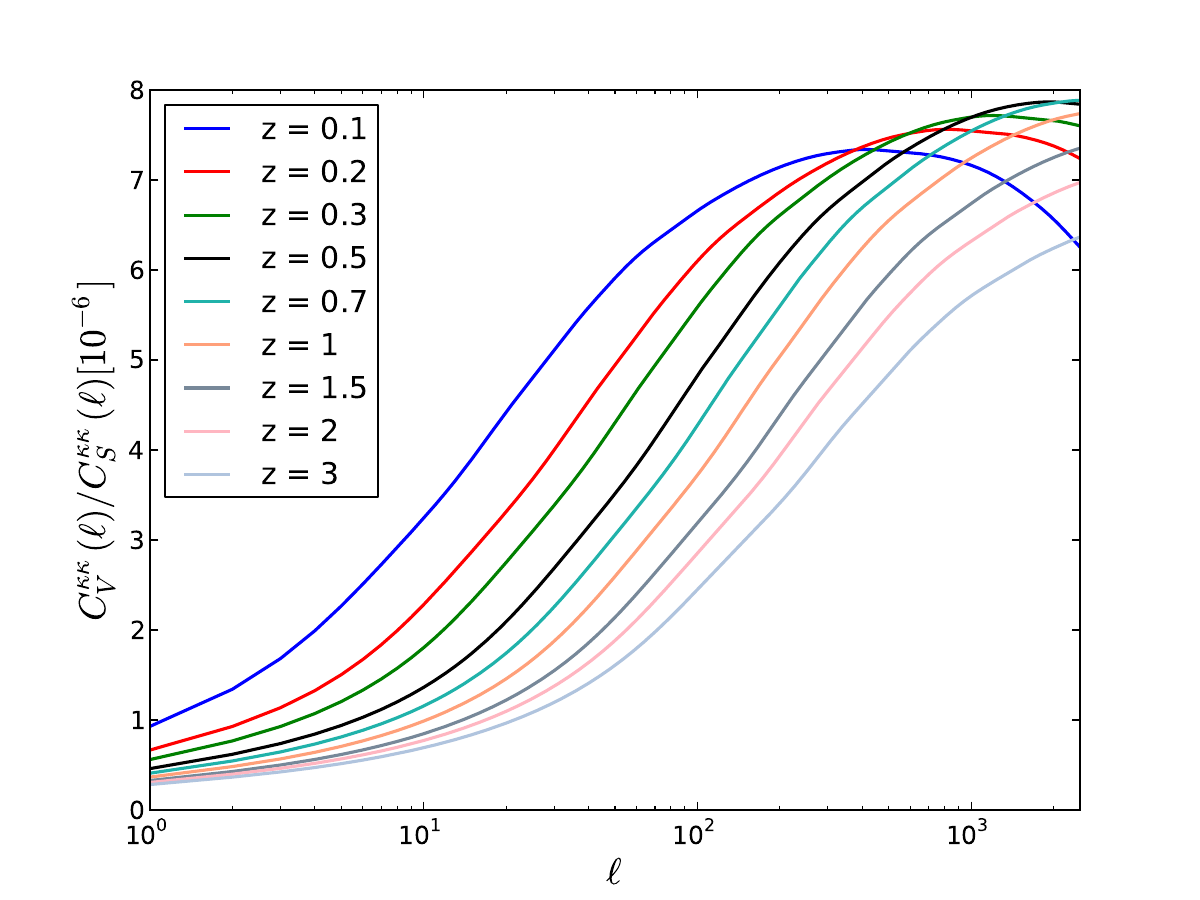}
\includegraphics[width=.66\columnwidth]{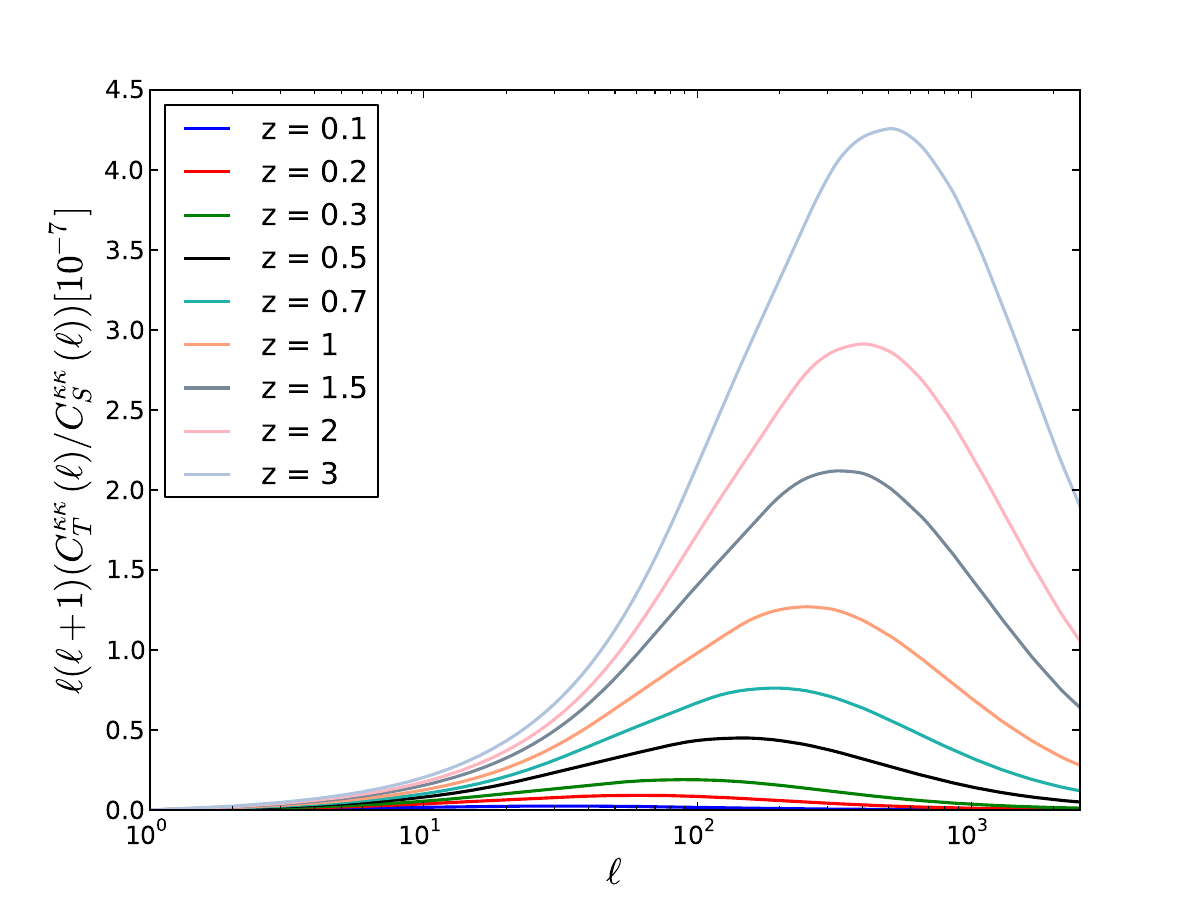}
\includegraphics[width=.66\columnwidth]{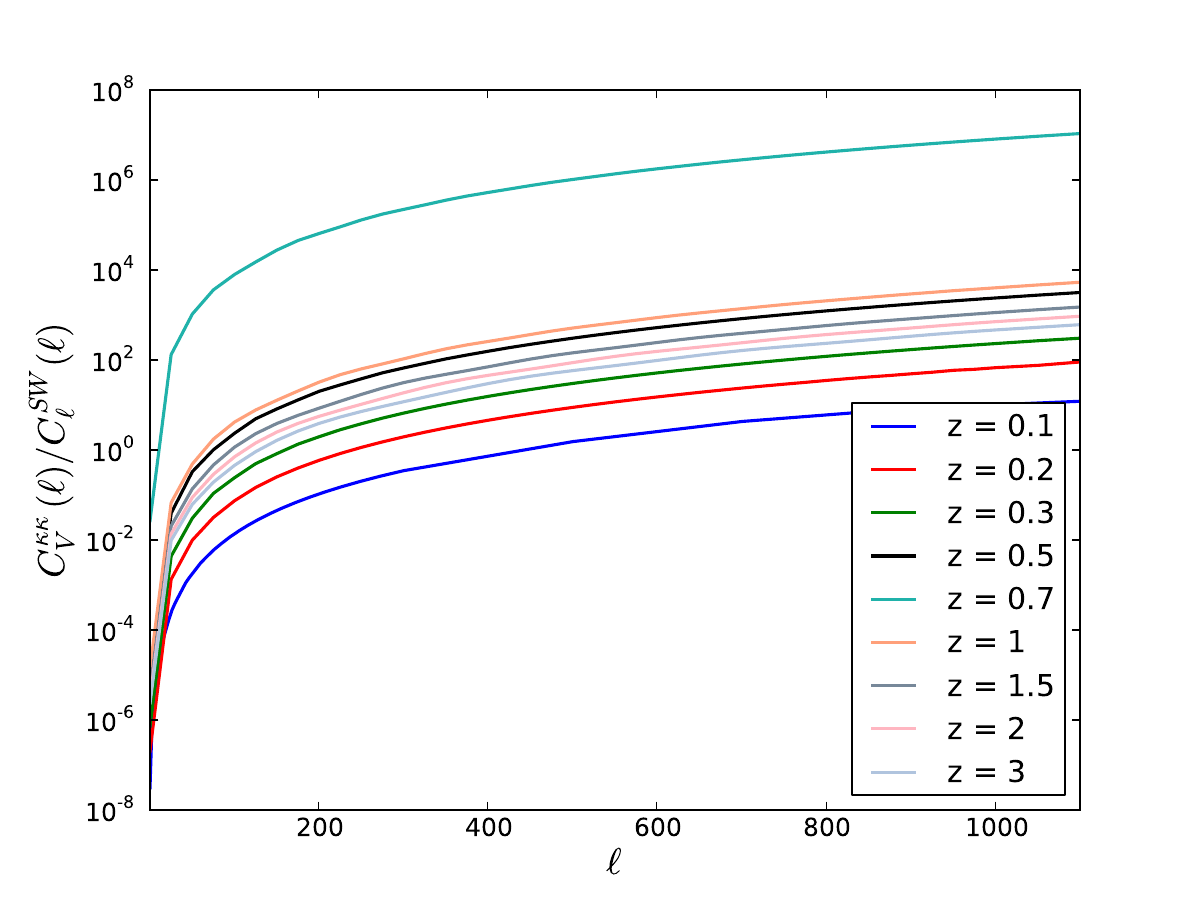}
\caption{The convergence contributions from the vector mode background (left), gravitational wave background (middle) which are both relative to the scalar contribution and the vector modes contribution relative to the Sachs-Wolfe term (right). There are all plotted for the same source distributions at single redshifts, using the distribution~(\ref{ch9t}).}
\label{points}
\end{center}
\end{figure*}

To start, let us assume that the sources are distributed at a single redshift so that
\be
n_s(\chi)=\delta(\chi-\chi_s)
\ee
which implies
\begin{equation}\label{ch9t}
 \hat g(\chi) = \frac{\chi_s-\chi}{\chi \chi_s}\Theta(\chi_s-\chi),
\end{equation}
where $\Theta$ is the Heaviside distribution.
This unrealistic but simple assumption provides a good way to understand the lensing effects as a function of redshift.  Fig.~\ref{points} depicts the contribution to the lensing spectra for shells with sources located at different redshifts normalised to the scalar contribution. As we can see, the relative contribution from the vector modes is largest at low redshift, reflecting the fact that vector modes continue to grow at late times. It can also be noticed that second order vector modes completely dominates the Sachs-Wolfe term at small scales (large $\ell$). The fact that SW term tends to zero at $z \simeq 0.7$ accounts for the large amplitude of the ratio $C^{\kappa \kappa}_{V}(\ell)/C^{SW}_{\ell}(\ell)$ at $z = 0.7$. Note that the difference in the pre-factors of both SW and Doppler terms explains the difference in redshifts at which each of them crosses zero. 

In order to obtain more realistic orders of magnitude, we consider source distributions similar to the one of the future Euclid and SKA experiments. The normalised Euclid redshift distribution has the form given in Refs.~\cite{Beynon:2009yd,Amendola:2012ys,euclid}:
\begin{equation}
n(z)=Az^{2}\exp\left[-\left(\frac{z}{z_{0}}\right)^{\beta}\right]
\end{equation}
with $A=5.792$, $\beta=1.5$ and $z_{0}=0.64$, which gives a median redshift $z_{m}\sim0.9$. 

For SKA we make use of the SKA Simulated Skies simulations~\cite{Wilman:2008ew}. These are simulations of the submillimeter radio source population. We use all the extragalactic radio continuum sources in the central $10\times10$ sq. degrees out to a redshift of $z=20$. In these simulations, the sources are drawn from either observed or extrapolated luminosity functions and grafted onto an underlying dark matter distribution with biases which reflect their measured large-scale clustering. We then construct a redshift distribution that we paramaterise as 
\begin{equation}
n(z)=A\frac{z^n}{(1+z)^m}\exp\left[-\frac{(a+bz)^{2}}{(1+z)^2}\right]
\end{equation}   
with best fit parameters  $a = -1.806, b = 0.388, m = 2.482, n = 0.838$ and $A=1.610$ and normalise the distribution at $z=20$, which gives a description accurate to the percent level, which is good enough for our purposes. Note that this redshift distribution represents the very best case scenario since all sources from the simulation have been used in its construction, and no further observational cuts were included.

\begin{figure*}
  \includegraphics[width=.65\columnwidth]{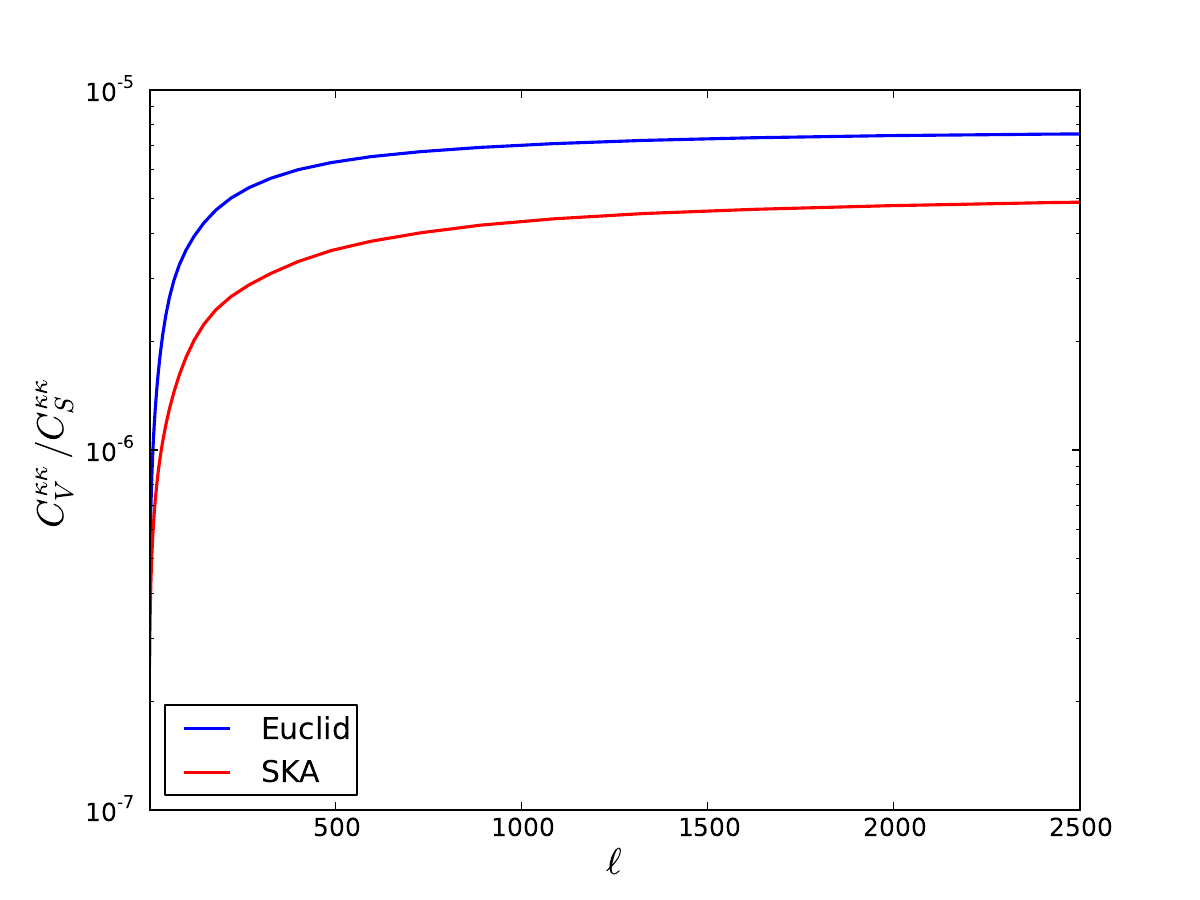} 
  \includegraphics[width=.65\columnwidth]{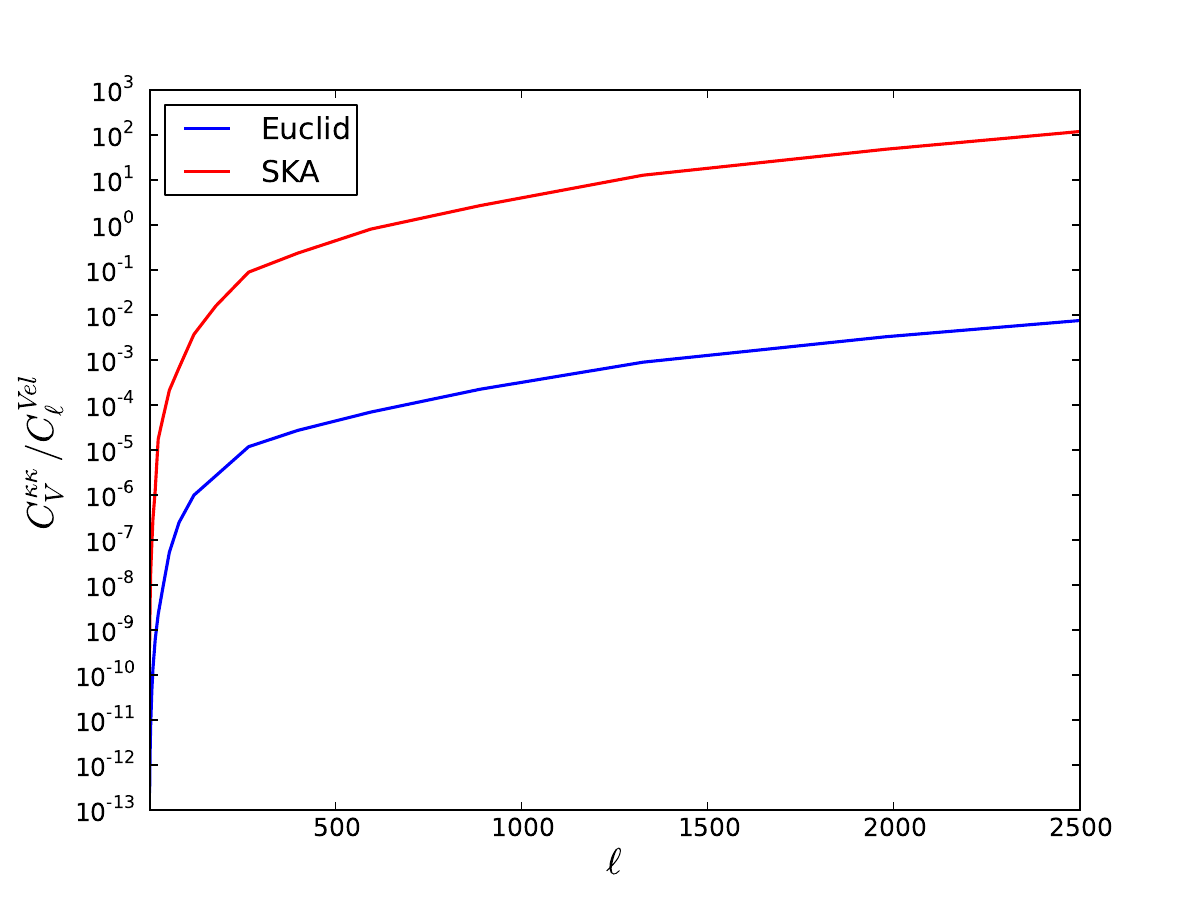} 
  \includegraphics[width=.65\columnwidth]{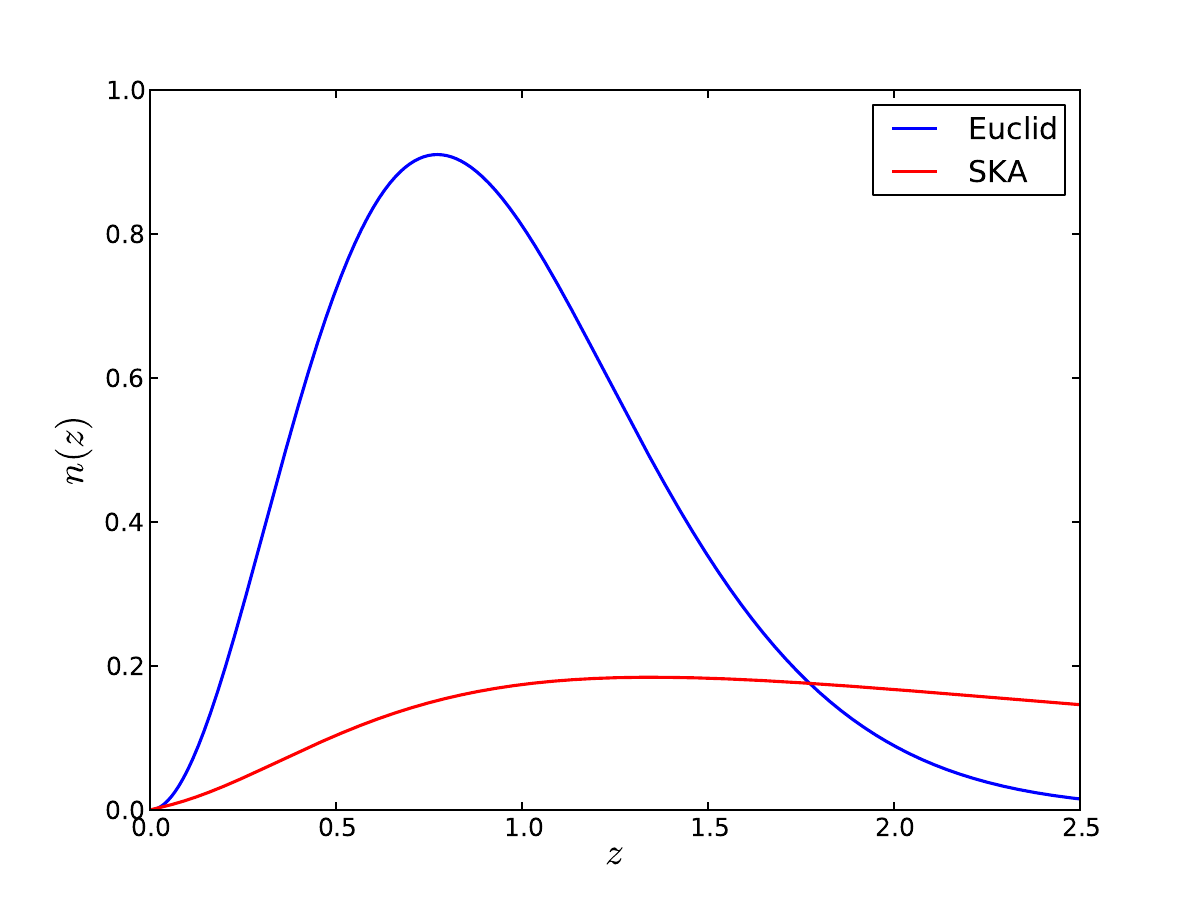}  
  \caption{\label{euclid-SKA}  Ratio between the angular power spectra of the vectors $C^{\kappa \kappa}_{V}$ to scalars $C^{\kappa \kappa}_{S}$ (left) and to the doppler term $C^{Vel}_{\ell}$ (middle left) and finally the ratio between the doppler term and scalars as a function of multipole $\ell$ for two surveys, blue line (Euclid-like), red line (SKA-like). The survey geometries are shown right.} 
\end{figure*}

These source distributions can be used to compute the vector convergence spectrum for both surveys. Fig.~\ref{euclid-SKA} compares its amplitude to the standard scalar contribution, showing that it is typically $10^{-5}$ times smaller. Whereas compared to the Doppler contribution, its amplitude is about $10^{2}$ larger and $10^{-2}$ smaller on small scales respectively for a SKA-like survey and for a Euclid-like survey~-- see Fig.~\ref{euclid-SKA}. Interestingly, the vector contribution is subdominant for Euclid, for which the main correction arises from the Doppler term, while for SKA-like geometry the vector contribution is typically 1-100 times larger than the Doppler one for $\ell>500$. On larger angular scales, the Doppler term always dominates~-- see Fig.~\ref{euclid-SKA}, where on large angular scales the Doppler term totally prevails over the scalar contribution by about 5 orders of magnitude.

\section{Conclusions}\label{sec5}

This article has evaluated the amplitude of relativistic contributions to the weak lensing power spectra. We have considered the gravitational wave and vector mode backgrounds which are sourced at second-order by density perturbations.  
The amplitude of these backgrounds are completely fixed once the normalisation of the scalar power spectrum in the linear regime is determined. As these are purely relativistic degrees of freedom they set the lower limit for all relativistic effects on cosmological modelling. While the gravitational wave background is very small in relation to the scalars, the vectors, which represent frame dragging in the metric, give corrections to the metric at nearly the percent level.
The effect of these contributions on weak lensing convergence predictions have been computed in order to understand if they can either be detected, or bias the analysis of future weak lensing experiments, such as Euclid or SKA. We have compared them to the usual gravitational lensing contribution, the two Sachs-Wolfe contributions as well as the Doppler lensing contribution~\cite{Bacon:2014uja}.

First, we have shown that even though the non-linear tensor mode background dominates over any possible primordial gravitational wave contribution, its effect on weak lensing is completely negligible, by 10 to 12 order of magnitudes (see Figs.~\ref{p12} and~\ref{points}).

Then, we have shown that the vector contribution to the convergence, while small, can dominate over the Doppler lensing at high redshift~-- but there it is swamped by gravitational lensing by density perturbations. We have shown this both for point sources and for two survey geometries. The vectors are actually more important than the Doppler term for SKA-like source distributions on small scales, but not for a Euclid like survey. For both of these surveys the vectors only reach about $10^{-3}$\% that of the normal gravitational lensing contribution, and so can be safely neglected. Nevertheless, it is interesting that the vector contribution can be as important as some linear terms. 

We have also recovered that although the frame dragging effect is small, it becomes more important than both ISW and SW above $\ell \ge 50$. This comes to corroborate the fact that for observations, neglecting the 2 first order Sachs-Wolfe terms is a good approximation.

In this analysis, the non-linear effects of the metric perturbations have been described at second order while weak lensing was described assuming that the Born approximation still holds. In principle, one needs also to take into account second order effects on the geodesic deviation equation~\cite{Seitz:1994xf, Cooray:2002mj,Dodelson:2005zj,Schaefer:2005up,Shapiro:2006em}, as fully described in Refs.~\cite{Bernardeau:2009bm,Bernardeau:2011tc}. 

There are a huge variety of second-order effects which come into the convergence. We have only considered two contributions which arise from non-linear dynamical effects which happen as structure forms. Many contributions appear when calculating the lensing convergence itself~\cite{Umeh:2012pn,Umeh:2014ana,BenDayan:2012wi}, and these also need to be analysed in a similar manner to that presented here to determine whether relativistic effects are important for future observations of magnification.

\section*{Acknowledgements}
SA, PP and OU are funded by the South African Square Kilometre Array Project. CC acknowledges funding from the NRF (South Africa).
This work made in the ILP LABEX (under reference ANR-10-LABX-63) was supported by French state funds managed by the ANR 
within the Investissements d'Avenir programme under reference ANR-11-IDEX-0004-02. JPU thanks the University of Cape Town for hospitality during the late stages of this project and Yannick Mellier for discussions.

\appendix\widetext
\section{Angular power spectrum of the scalar modes}\label{app0}

We follow the standard description of weak lensing in a full sky analysis, following e.g, Refs.~\cite{pubook,ehlers,bartelman} and refer to Refs.~\cite{ppu,Yamauchi:2013fra} for more recent developments of the formalism.

Taking into account that one can neglect the anisotropic stress, the deflecting potential integrated over the line of sight~(\ref{eq39}) reduces to
\be
 \psi(\bn)=2\int_0^\infty\dd\chi\,\, \hat g(\chi)\Phi[\bx(\bn),\eta]\, ,
\ee
where $\hat g$ is defined in Eq.~(\ref{ch81}). By inserting the Fourier decomposition Eq.~(\ref{e.Fdec}) and expanding the exponential in spherical harmonics as
\begin{equation}\label{a4}
 \exp({\rm i}\bk\cdot\bx)=4\pi\sum_{\ell m}{\rm i}^\ell j_\ell(kx) Y_{\ell m}(\hat\bk)Y_{\ell m}(\hat\bx)\,,
\end{equation}
where the $j_\ell$ are the spherical Bessel functions, the components $\psi_{\ell m}$ are given by
\begin{eqnarray}
 \psi_{\ell m} &=&  \frac{4\,{\rm i}^\ell}{\sqrt{2\pi}}\int_0^\infty\dd\chi\int {\dd^3\bk}\,\, \hat g(\chi)  
 j_\ell(k\chi)\Phi(\bk,\chi)Y_{\ell m}(\hat\bk)\,,
\end{eqnarray}
where we have replaced $\eta=\eta_0-\chi$ by $\chi$ since the integral is evaluated on the past lightcone. It follows that
\begin{eqnarray}\label{eA4}
 \langle\psi_{\ell m}\psi_{\ell'm'}^*\rangle
   &=& 16\pi \int_0^\infty \dd\chi\int_0^\infty\dd\chi' \int_0^\infty\frac{\dd k}{k} \,\,
   \hat g(\chi)\hat g(\chi')
   j_\ell(k\chi)j_\ell(k\chi')\mathcal{P}_\Phi(k,\chi,\chi')
\delta_{\ell\ell'}\delta_{mm'}
   \,,
\end{eqnarray} 
using Eq.~(\ref{e.powerS}), integrating over $\bk'$, then decomposing $\dd^3\bk=k^2\dd k\dd^2\hat\bk$ and integrating the product of spherical harmonics over $\hat\bk$ to get the term $\delta_{\ell\ell'}\delta_{mm'}$. The expressions for the scalar $C_\ell$'s in the text follow directly.

\section{Angular power spectrum of the Doppler term}\label{appA}

Starting from the expression~(\ref{jeqv}) for the convergence associated to the Doppler effect in which $v_i$ is given by Eq.~(\ref{j2.2}), and using the decomposition of the gravitational potential described in \S~\ref{sec2-1}, one obtains that
\begin{equation}
v_i = -\frac{2a(\eta)}{3\Omega_{\rm m}H_{0}^{2}}[g(\eta)'+\mathcal{H}g(\eta)]\partial_i\Phi.
\end{equation}
Decomposing the gravitational potential in Fourier mode as in Eq.~(\ref{e.Fdec}), with the definition of its power spectrum given in Eq.~(\ref{e.powerS}), one gets
\begin{equation}\label{e2}
\kappa(\bn) = A \int_{0}^{\infty}\dd\chi n_{s}(\chi)a(\chi)\left(1-\frac{1}{\chi \mathcal{H}(\chi)}\right)\int \frac{\dd^3\bk}{(2\pi)^{3/2}}\Phi(\bk,\eta)n^i\partial_i\left(e^{{\rm i}\bk.\bx}\right)  
\end{equation}
where the coefficient $A$ is given by
$$
A = \frac{2}{3\Omega_{\rm m}H_{0}^{2}}. 
$$
Now, using that
\begin{equation}
n^i\partial_i\left( \exp({\rm i}\bk\cdot\bx)\right)=4\pi\sum_{\ell m}{\rm i}^\ell kj'_\ell(k\chi)
 Y_{\ell m}(\hat\bk)Y_{\ell m}(\bn)\ ,
\end{equation}
with a prime on the spherical Bessel function denoting the derivative with respect to its argument, Eq.~(\ref{e2}) becomes
\begin{equation}
\kappa(\bn) = 4\pi A \int_{0}^{\infty}\dd\chi n_{s}(\chi)a(\chi)\left(1-\frac{1}{\chi \mathcal{H}(\chi)}\right)\int \frac{\dd^3\bk}{(2\pi)^{3/2}}\Phi(\bk,\eta)\sum_{\ell m}{\rm i}^\ell kj'_\ell(k\chi)Y_{\ell m}(\hat\bk)Y_{\ell m}(\bn)
\end{equation}
from which we can extract the components
\begin{equation}
\kappa_{\ell m} = 4\pi A \int_{0}^{\infty}\dd\chi n_{s}(\chi)a(\chi)\left(1-\frac{1}{\chi \mathcal{H}(\chi)}\right)\int \frac{\dd^3\bk}{(2\pi)^{3/2}}\Phi(\bk,\eta){\rm i}^\ell kj'_\ell(k\chi)Y_{\ell m}(\hat\bk).
\end{equation}
Its correlator is then given by
\begin{equation}\label{eB6}
\langle \kappa_{\ell m}\kappa_{\ell' m'}^*\rangle =  4\pi A^2 \int_0^\infty \dd\chi F(\chi)j'_{\ell}(k\chi)\int_0^\infty \dd\chi' F(\chi')j'_{\ell}(k\chi')\int_0^\infty \frac{\dd k}{k} \mathcal{P}_{v}(k,\chi,\chi')\delta_{\ell\ell'}\delta_{mm'}
\end{equation}
with
\be
 \mathcal{P}_{v}(k,\chi,\chi') \equiv k^2 \mathcal{P}_{\Phi}(k,\chi,\chi')
\ee
and where 
\be
F(\chi) \equiv n_s(\chi)a(\chi)\left(1-\frac{1}{\chi \mathcal{H}(\chi)}\right).
\ee
We finally get the formula of the angular power spectrum convergence associated to the Doppler contribution as
\begin{equation}\label{e6}
C^{v}_{\ell} = 4\pi A^2 \int_0^\infty \dd\chi F(\chi)\int_0^\infty \dd\chi' F(\chi')\int_0^\infty \frac{\dd k}{k} \mathcal{P}_{v}(k,\chi,\chi')j'_{\ell}(k\chi)j'_{\ell}(k\chi')\, .
\end{equation}
Since $\mathcal{P}_{\Phi}(k,\chi,\chi') = \mathcal{P}_{\Phi_i}(k)\tilde T(k,\chi)\tilde T(k,\chi')$ with
$$
\tilde T(k,\chi) = T(k)\left[g'(\chi)+\mathcal{H}g(\chi)\right]
$$
the angular spectrum reduces to
\begin{equation}\label{e15}
C^{v}_{\ell} = 4\pi A^2\int_0^\infty \frac{\dd k}{k} \mathcal{P}_{v_i}(k)\left[\int_0^\infty \dd\chi F(\chi)j'_{\ell}(k\chi)\tilde T(k,\chi)\right]^2\,.
\end{equation}

\section{Angular power spectrum of the Integrated Sachs-Wolfe term}\label{appD}
As discussed in the text, the Integrated Sachs-Wolfe also contribute to the cosmic convergence at first order
\be
\kappa_{\rm isw}(\bn)=2\int_0^{\infty} d\chi\ \hat g_{\rm isw1}(\chi)\Phi'(\bn,\chi) \\
+2\int_0^{\infty} d\chi \hat g_{\rm isw2}(\chi)\Phi(\bn,\chi)
\ee
with both $ \hat g_{\rm isw2}$ and $ \hat g_{\rm isw1}$ defined in the text. The harmonic expansions of both the first and the second terms, which we call $\kappa_{\rm isw1}$ and $\kappa_{\rm isw2}$ respectively give
\begin{equation}
\kappa_{\rm isw1}(\bn) = 8\pi \int_{0}^{\infty}\dd\chi \hat g_{\rm isw1}(\chi)\int \frac{\dd^3\bk}{(2\pi)^{3/2}}\Phi(\bk,\eta)\sum_{\ell m}{\rm i}^\ell j_\ell(k\chi)Y_{\ell m}(\hat\bk)Y_{\ell m}(\bn)
\end{equation}
for now we drop the $'$ which denotes the derivative of the potential with respect to the conformal time as it is taken into account by the time evolution of the transfer function. The second term 
\begin{equation}
\kappa_{\rm isw2}(\bn) = -8\pi \int_{0}^{\infty}\dd\chi \hat g_{\rm isw2}(\chi)\int \frac{\dd^3\bk}{(2\pi)^{3/2}}\Phi(\bk,\eta)\sum_{\ell m}{\rm i}^\ell j_\ell(k\chi)Y_{\ell m}(\hat\bk)Y_{\ell m}(\bn).
\end{equation}
It follows that the correlator contains three terms
\be
 \langle\kappa^{\rm isw}_{\ell m}\kappa^{\rm isw^*}_{\ell' m'}\rangle =  \langle\kappa^{\rm isw1}_{\ell m}\kappa^{\rm isw1^*}_{\ell' m'}\rangle +  \langle\kappa^{\rm isw2}_{\ell m}\kappa^{\rm isw2^*}_{\ell' m'}\rangle +  2\langle\kappa^{\rm isw1}_{\ell m}\kappa^{\rm isw2^*}_{\ell' m'}\rangle
 \ee
 thus
 \bea \label{iswE1}
 \langle\kappa^{\rm isw1}_{\ell m}\kappa^{\rm isw1^*}_{\ell' m'}\rangle 
 &=& 16\pi \int_0^\infty \dd\chi\int_0^\infty\dd\chi' \int_0^\infty\frac{\dd k}{k} \,\,
   \hat g_{\rm isw1}(\chi)\hat g_{\rm isw1}(\chi')
   j_\ell(k\chi)j_\ell(k\chi')\mathcal{P}_\Phi(k,\chi,\chi')
\delta_{\ell\ell'}\delta_{mm'}
 \eea
where $$\mathcal{P}_\Phi(k,\chi,\chi') = \mathcal{P}_{\Phi_i}(k)T_{\rm isw1}(k,\chi)T_{\rm isw1}(k,\chi')$$ and $$T_{\rm isw1}(k,\chi) = T(k)g'(\chi)$$ $g'(\chi)$ being the derivative of the growth suppression factor with respect to conformal time $\eta$. The second term that constitutes to the correlator
 \bea \label{iswE2}
 \langle\kappa^{\rm isw2}_{\ell m}\kappa^{\rm isw2^*}_{\ell' m'}\rangle 
 &=& 16\pi \int_0^\infty \dd\chi\int_0^\infty\dd\chi' \int_0^\infty\frac{\dd k}{k} \,\,
   \hat g_{\rm isw2}(\chi)\hat g_{\rm isw2}(\chi')
   j_\ell(k\chi)j_\ell(k\chi')\mathcal{P}_\Phi(k,\chi,\chi')
\delta_{\ell\ell'}\delta_{mm'}
 \eea
with $$\mathcal{P}_\Phi(k,\chi,\chi') = \mathcal{P}_{\Phi_i}(k)T_{\rm isw2}(k,\chi)T_{\rm isw2}(k,\chi')$$ and $$T_{\rm isw2}(k,\chi) = T(k)g(\chi).$$And the last term yields
 \bea \label{iswE3}
 \langle\kappa^{\rm isw1}_{\ell m}\kappa^{\rm isw2^*}_{\ell' m'}\rangle 
 &=& -16\pi \int_0^\infty \dd\chi\int_0^\infty\dd\chi' \int_0^\infty\frac{\dd k}{k} \,\,
   \hat g_{\rm isw1}(\chi)\hat g_{\rm isw2}(\chi')
   j_\ell(k\chi)j_\ell(k\chi')\mathcal{P}_\Phi(k,\chi,\chi')
\delta_{\ell\ell'}\delta_{mm'}
 \eea
letting $$\mathcal{P}_\Phi(k,\chi,\chi') = \mathcal{P}_{\Phi_i}(k)T_{\rm isw1}(k,\chi)T_{\rm isw2}(k,\chi').$$ So, the total contribution of the Integrated Sachs-Wolfe term ($C^{\rm isw}_{\ell}$) to the convergence is thus given by the sum of each $C_{\ell}'s$ extracted from each of the terms that composes the correlator i.e
\be
C^{\rm isw}_{\ell} = C^{\rm isw1,\rm isw1}_{\ell} + C^{\rm isw2,\rm isw2}_{\ell} - 2C^{\rm isw1,\rm isw2}_{\ell}.
\ee

\section{Angular power spectrum of the Sachs-Wolfe term}\label{appE}
The contribution to the convergence of the Sachs-Wolfe term reads
\be
\kappa_{\rm sw}(\bn)=\int_0^{\infty}d\chi n_{s}(\chi)\left(2-\frac{1}{\mathcal{H}\chi}\right)\Phi(\bn,\chi).
\ee
Using the Fourier decomposition of the potential and expanding the plane waves we arrive at
\begin{equation}
\kappa_{\rm sw}(\bn) = 4\pi \int_{0}^{\infty}\dd\chi n_{s}(\chi)\left(2-\frac{1}{\chi \mathcal{H}(\chi)}\right)\int \frac{\dd^3\bk}{(2\pi)^{3/2}}\Phi(\bk,\eta)\sum_{\ell m}{\rm i}^\ell j_\ell(k\chi)Y_{\ell m}(\hat\bk)Y_{\ell m}(\bn)
\end{equation}
and the coefficients $\kappa^{\rm sw}_{\ell m}$ are given by
\begin{equation}
\kappa^{\rm sw}_{\ell m} = 4\pi \int_{0}^{\infty}\dd\chi n_{s}(\chi)\left(2-\frac{1}{\chi \mathcal{H}(\chi)}\right)\int \frac{\dd^3\bk}{(2\pi)^{3/2}}\Phi(\bk,\eta){\rm i}^\ell j_\ell(k\chi)Y_{\ell m}(\hat\bk)
\end{equation}
so that
\begin{eqnarray}\label{swD1}
 \langle\kappa^{\rm sw}_{\ell m}\kappa^{\rm sw^*}_{\ell' m'}\rangle
   &=& 4\pi \int_0^\infty \dd\chi\int_0^\infty\dd\chi' \int_0^\infty\frac{\dd k}{k} \,\,
   \hat g_{\rm sw}(\chi)\hat g_{\rm sw}(\chi')
   j_\ell(k\chi)j_\ell(k\chi')\mathcal{P}_\Phi(k,\chi,\chi')
\delta_{\ell\ell'}\delta_{mm'}
   \
\end{eqnarray} 
where we define $$\hat g_{\rm sw}(\chi) = n_{s}(\chi)\left(2-\frac{1}{\mathcal{H}\chi}\right).$$
Extracting $C^{\rm sw}_{\ell}$ from (\ref{swD1}) is straightforward
\section{Angular power spectrum of the vector modes}\label{appB}

The lensing potential integrated along the line of sight associated with the vector modes is given by
$$
 \psi(\bn)=\int_0^\infty \dd\chi \hat g(\chi)n^iV_i[\bx(\bn),\chi].
$$
We decompose the vector perturbations in Fourier modes as in Eq.~(\ref{eFvec}). The polarisation vectors can be expressed as
$$
 \bm{e}^\pm =\frac{1}{\sqrt{2}}(\bm{e}_1\pm {\rm i} \bm{e}_2)\ ,
$$
so that
\begin{equation}\label{eq11}
 n^ie^\pm_i =\frac{1}{\sqrt{2}}\sin\theta\hbox{e}^{\pm  {\rm i}\varphi}\ .
\end{equation}

The power spectrum of each polarisation is then defined as
\begin{equation}\label{e.pB}
 \langle V_a(\bk,\eta)V_b^*(\bk',\eta')\rangle=\frac{2\pi^2}{k^3}\mathcal{P}_V(k,\eta,\eta')
 \delta^{(3)}(\bk-\bk')\delta_{ab}\ ,
\end{equation}
assuming that the two polarisations are independent and using local isotropy to deduce that they enjoy the same spectrum. It follows that
\begin{equation}
 \psi(\bm{n}) = \int_0^{\infty}\dd\chi\hat g(\chi)\sum_{\lambda=\pm} \int\frac{\dd^3\bk}{(2\pi)^{3/2}}
 V_\lambda(\bk,\eta)
 \frac{1}{\sqrt{2}}\sin\theta\hbox{e}^{\lambda {\rm i} \varphi}
 \hbox{e}^{{\rm i}\bk\cdot\bx}\ .
\end{equation}
Contrary to the scalar case, we cannot simply decompose the exponential to read $\psi_{\ell m}$ because of the extra geometric factor. The simplest is to extract it as
\be\label{esph}
 \psi_{\ell m } = \int \psi(\bm{n})Y_{\ell
 m}^*(\bm{n})\dd^2\bm{n}\ .
\ee 
so that is given by
\begin{equation}
 \psi_{\ell m } = \frac{1}{\sqrt{2}}\int_0^{\infty}\dd\chi
  \hat g(\chi)\sum_{\lambda=\pm} \int\frac{\dd^3\bk}{(2\pi)^{3/2}}
 V_\lambda(\bk,\eta) \int\sin\theta\hbox{e}^{\lambda {\rm i}\varphi}
 Y_{\ell m}^*(\bm{n})\hbox{e}^{{\rm i}\bk\cdot\bx}\dd^2\bm{n}\ .
\end{equation}
Now, using that
$$
 \sin\theta\hbox{e}^{\lambda {\rm i}\varphi} =
 -2\lambda\sqrt{\frac{2\pi}{3}}Y_{1\lambda}(\bn)\ 
$$
and decomposing the exponential in spherical harmonics, one gets
\begin{equation}
 \psi_{\ell m } = 4\pi\sqrt{2} \int_0^{\infty}\dd\chi
  \hat g(\chi)\sum_{\lambda=\pm} (-\lambda)\sqrt{\frac{2\pi}{3}}
   \int\frac{\dd^3\bk}{(2\pi)^{3/2}}V_\lambda(\bk,\eta) 
  \sum_{LM} {\rm i}^Lj_L(k\chi)Y_{LM}(\hat\bk) \mathcal{A}^\lambda_{LM,\ell m}
 \ .
\end{equation}
The integral over the 3 spherical harmonics,
\begin{eqnarray}
 \mathcal{A}^\lambda_{LM,\ell m} &=&
 \int Y_{1\lambda}(\bn)Y_{LM}(\bn) Y_{\ell m}^*(\bm{n})\dd^2\bm{n}
 \ ,
\end{eqnarray}
is conveniently computed by first assuming that $\hat\bk$ is along the $z$-axis so that $Y_{LM}(\hat\bk)=\sqrt{(2L+1)/4\pi}\delta_{M0}$. We thus need to
evaluate $\mathcal{A}^\lambda_{L0,\ell m}$, which is only non-vanishing when
$m=\lambda\quad L=\ell\pm1$ so that the only non-vanishing coefficients are
\begin{eqnarray}
 \mathcal{A}^{1\pm1}_{\ell+1 0,\ell \pm1} =
  -\frac{1}{2}\sqrt{\frac{3}{2\pi}}\sqrt{\frac{\ell(\ell+1)}{(2\ell+1)(2\ell+3)}}
  , \qquad
 \mathcal{A}^{1\pm1}_{\ell-1 0,\ell \pm1} =
  \frac{1}{2}\sqrt{\frac{3}{2\pi}}\sqrt{\frac{\ell(\ell+1)}{(2\ell+1)(2\ell-1)}}
  .
\end{eqnarray}
The sum
\begin{equation}
 \alpha^\lambda_{\ell m}(\bk=k\bm{e}_z)\equiv\sum_{LM} {\rm i}^Lj_L(k\chi)Y_{LM}(\hat\bk) \mathcal{A}^\lambda_{LM,\ell m}
 \end{equation}
 then reduces to
 \begin{equation}
 \alpha^\lambda_{\ell m}(\bk=k\bm{e}_z)= {\rm i}^{\ell-1}j_{\ell-1}\sqrt{\frac{2\ell-1}{4\pi}}  \mathcal{A}^{1\lambda}_{\ell-1 0,\ell \pm1} - {\rm i}^{\ell+1}j_{\ell+1}\sqrt{\frac{2\ell+3}{4\pi}}\mathcal{A}^{1\lambda}_{\ell+1 0,\ell \pm1},
 \end{equation} 
and, after gathering the Bessel functions,
\begin{equation}
 \alpha^\lambda_{\ell m}(\bk=k\bm{e}_z)=-
  {\rm i}^{\ell+1}\frac{1}{\sqrt{2}}\sqrt{\frac{3}{2\pi}}\sqrt{\frac{2\ell+1}{4\pi}}j_\ell^{(11)}(k\chi)
  \delta_{m\lambda}
 \  
\end{equation}
with
$$
 j_\ell^{(11)}(x)\equiv\sqrt{\frac{\ell(\ell+1)}{2}}\frac{j_\ell(x)}{x} \, .
$$
Now, to evaluate the same quantity for any $\hat\bk$, we need to perform a rotation $R(\hat\bk)$ that brings the $\bm{e}_z$ along $\hat\bk$. Under such a rotation,
\begin{equation}
 \alpha^\lambda_{\ell m}(\bk)=
 \sum_{m'=\pm1} D_{m,m'}^\ell[R(\hat\bk)]\alpha^\lambda_{\ell m'}(k\bm{e}_z)
 \ ,
\end{equation}
where
\begin{equation}\label{DDint}
 \int\dd\hat\bk
 D_{m,\pm1}^\ell[R(\hat\bk)]\left(D_{m',\pm1}^{\ell'}[R(\hat\bk)]\right)^*=
 \frac{4\pi}{2\ell+1}\delta_{\ell\ell'}\delta_{mm'}\ .
\end{equation}
So, finally, we have
\begin{equation}
 \psi_{\ell m } = 4\pi\int_0^{\infty}\dd\chi \hat g(\chi)
   {\rm i}^{\ell+1}\sqrt{\frac{2\ell+1}{4\pi}}j_\ell^{(11)}(k\chi) \sum_{\lambda=\pm} \lambda
   \int\frac{\dd^3\bk}{(2\pi)^{3/2}}V_\lambda(\bk,\eta)\sum_{a=\pm1} D_{m,a}^\ell[R(\hat\bk)]
 \ .
\end{equation}
Using Eq.~(\ref{e.pB}) and integrating over $\hat\bk$ while exploiting relation (\ref{DDint}), it follows that
\begin{eqnarray}\label{eC14}
  \langle\psi_{\ell m}\psi_{\ell'm'}^*\rangle
   &=& 4\pi\int_0^\infty \dd\chi \int_0^\infty \dd\chi'\int\frac{\dd k}{k}
   \mathcal{P}_V (k,\eta,\eta')\hat g(\chi)\hat g(\chi')
   j_\ell^{(11)}(k\chi)j_\ell^{(11)}(k\chi')
   \delta_{\ell\ell'}\delta_{mm'}\  .
\end{eqnarray}

\section{Angular power spectrum of the tensor modes}\label{appC}

The same method as for vectors can be followed for tensor modes. The potential integrated along the line of sight is now given by
\begin{equation}
 \psi(\bn)=\int_0^\infty \hat g(\chi)n^in^j h_{ij}[\bx(\bn),\chi]\dd\chi\ .
\end{equation}
We decompose the tensor perturbations in Fourier modes as in Eq.~(\ref{eFtens}) in which the polarization tensor is explicitely given by
$$
\varepsilon_{ij}^\lambda = \frac{e_{i}^{1}e_{j}^{1}\delta_{+}^{\lambda} +e_{i}^{2}e_{j}^{2}\delta_{-}^{\lambda} }{\sqrt{2}}
$$
so that
$$
n^{i}n^{j}\varepsilon_{ij}^{\pm} = \frac{1}{2\sqrt{2}}(\sin\theta)^{2}\hbox{e}^{\pm2 {\rm i}\varphi} .
$$
The power spectrum of the two polarisations is defined as
\begin{equation}\label{e.pE}
 \langle h_{a}(\bk,\eta)h_b^*(\bk',\eta')\rangle=\frac{2\pi^2}{k^3}\mathcal{P}_T(k,\eta,\eta')
 \delta^{(3)}(\bk-\bk')\delta_{ab}\ .
\end{equation}
It follows that
\begin{equation}
 \psi(\bm{n}) = \int_0^\infty  \dd\chi\hat g(\chi)\sum_{\lambda=\pm} \int\frac{\dd^3\bk}{(2\pi)^{3/2}}
 h_\lambda(\bk,\eta)
 \frac{1}{2\sqrt{2}}(\sin\theta)^{2}\hbox{e}^{\lambda2 {\rm i}\varphi}
 \hbox{e}^{ {\rm i}\bk\cdot\bx}\ .
\end{equation}
Setting $2\lambda = \gamma$ and using
$$
(\sin\theta)^{2}\hbox{e}^{\gamma  {\rm i}\varphi} = 4\sqrt{\frac{2\pi}{15}}Y_{2\gamma}(\bn),
$$
the expression of the coefficients $\psi_{\ell m}$ are obtained from Eq.~(\ref{esph}) as
\begin{equation}
 \psi_{\ell m } = 4\pi\sqrt{2}\int_0^\infty \dd\chi
  \hat g(\chi) \sum_{\gamma=\pm2}\sqrt{\frac{2\pi}{15}}
   \int\frac{\dd^3\bk}{(2\pi)^{3/2}} h_\gamma(\bk,\eta)
  \sum_{LM}  {\rm i}^Lj_L(k\chi)Y_{LM}(\hat\bk) \mathcal{A}^\gamma_{LM,\ell m}
 \ 
\end{equation}
with
\begin{eqnarray}
 \mathcal{A}^\gamma_{LM,\ell m} &=&
 \int Y_{2\gamma}(\bn)Y_{LM}(\bn) Y_{\ell m}^*(\bm{n})\dd^2\bm{n}
 \ .
\end{eqnarray}
By pure analogy with the previous case, it is convenient to first calculate this assuming that $\hat\bk$ is along the $z$-axis so that $Y_{LM}(\hat\bk)=\sqrt{(2L+1)/4\pi}\delta_{M0}$. We then need to evaluate $ \mathcal{A}^\gamma_{LM,\ell m}$ which is only non-vanishing when $m=\gamma \quad L=\ell\pm2,\ell$ so that the non-zero coefficients are 
\begin{eqnarray}
 \mathcal{A}^{2\pm2}_{\ell 0,\ell \pm2} &=&
  -\frac{1}{2}\sqrt{\frac{15}{2\pi}}\sqrt{\frac{(\ell+2)!}{(\ell-2)!}}\frac{1}{(2\ell+3)(2\ell-1)}\
  , \\
 \mathcal{A}^{2\pm2}_{\ell+2 0,\ell \pm2} &=&
  \frac{1}{4}\sqrt{\frac{15}{2\pi}}\sqrt{\frac{(\ell+2)!}{(\ell-2)!}}\frac{1}{\sqrt{(2\ell+1)(2\ell+5)}(2\ell+3)}\
  , \\
  \mathcal{A}^{2\pm2}_{\ell-2 0,\ell \pm2} &=&
  \frac{1}{4}\sqrt{\frac{15}{2\pi}}\sqrt{\frac{(\ell+2)!}{(\ell-2)!}}\frac{1}{\sqrt{(2\ell-3)(2\ell+1)}(2\ell-1)}\
  .
\end{eqnarray}
The sum
\begin{equation}
\tilde{\alpha}^\gamma_{\ell m}(\bk=k\bm{e}_z)\equiv\sum_{LM}  {\rm i}^Lj_L(k\chi)Y_{LM}(\hat\bk) \mathcal{A}^\gamma_{LM,\ell m}
 \end{equation}
then reduces to 3 terms as
 \begin{equation}
\tilde{\alpha}^\gamma_{\ell m}(\bk=k\bm{e}_z) =  {\rm i}^\ell j_{\ell}(k\chi)Y_{\ell0}(\hat\bk) \mathcal{A}^{2\pm2}_{\ell 0,\ell \pm2}+ {\rm i}^{\ell+2} j_{\ell+2}(k\chi)Y_{\ell+20}(\hat\bk) \mathcal{A}^{2\pm2}_{\ell+2 0,\ell \pm2}+ {\rm i}^{\ell-2} j_{\ell-2}(k\chi)Y_{\ell-20}(\hat\bk) \mathcal{A}^{2\pm2}_{\ell-2 0,\ell \pm2}.
 \end{equation}
After simplifying and gathering the Bessel functions, it gives
\begin{equation}
\tilde{\alpha}^\gamma_{\ell m}(\bk=k\bm{e}_z) = \frac{1}{\sqrt{2}} {\rm i}^{\ell+2}\sqrt{\frac{15}{2\pi}}\sqrt{\frac{2\ell+1}{4\pi}}j_{\ell}^{(22)}(k\chi)\delta_{\gamma m}
\end{equation}
with
$$
j_{\ell}^{(22)}(x) \equiv \sqrt{\frac{1}{8}\frac{(\ell+2)!}{(\ell-2)!}}\frac{j_{\ell}(x)}{x^{2}}.
$$
To finish, we need to evaluate the same quantity for any $\hat\bk$ by performing a rotation $R(\hat\bk)$ that brings the $\bm{e}_z$ along $\hat\bk$. Under such a rotation,
\begin{equation}
\tilde{ \alpha}^\lambda_{\ell m}(\bk)=
 \sum_{m'=\pm2} D_{m,m'}^\ell[R(\hat\bk)]\tilde{\alpha}^\lambda_{\ell m'}(k\bm{e}_z)
 \ ,
\end{equation}
where the orthogonality relation of the Wigner $D$-functions,
\begin{equation}\label{DD2}
 \int\dd\hat\bk
 D_{m,\pm2}^\ell[R(\hat\bk)]\left(D_{m',\pm2}^{\ell'}[R(\hat\bk)]\right)^*=
 \frac{4\pi}{2\ell+1}\delta_{\ell\ell'}\delta_{mm'}\ ,
\end{equation}
is used to get the coefficients of the expansion  as
\begin{equation}
 \psi_{\ell m } = 4\pi
  \sqrt{\frac{2\ell+1}{4\pi}}\int_0^\infty \dd\chi
  \hat g(\chi)\sum_{\gamma=\pm2}
   \int\frac{\dd^3\bk}{(2\pi)^{3/2}} h_\gamma(\bk,\eta)  {\rm i}^{\ell+2}j_{\ell}^{(22)}(k\chi) \sum_{a=\pm2} D_{m,a}^\ell[R(\hat\bk)]\ .
\end{equation}
Using Eq.~(\ref{e.pE}) and integrating over $\hat\bk$ while exploiting relation (\ref{DD2}) leads to
\begin{eqnarray}\label{eD15}
  \langle\psi_{\ell m}\psi_{\ell'm'}^*\rangle
   &=& 4\pi\int_0^{\infty}\dd\chi\int_0^{\infty}\dd\chi'\int\frac{\dd k}{k}
   \mathcal{P}_T(k,\eta,\eta')\hat g(\chi)\hat g(\chi')
   j_\ell^{(22)}(k\chi)j_\ell^{(22)}(k\chi')
   \delta_{\ell\ell'}\delta_{mm'}\  .
\end{eqnarray}

\end{document}